\documentclass[aps,
twocolumn,
showpacs,preprintnumbers,amsmath,amssymb,superscriptaddress,prb,floatfix,10pt]{revtex4-2}

\usepackage{graphicx}
\usepackage{dcolumn}
\usepackage{bm}
\usepackage{graphics}
\usepackage{subfigure}
\usepackage{epsfig}
\usepackage{epstopdf}
\usepackage{xcolor}
\usepackage{multirow}
\usepackage{mathtools}
\usepackage{comment}
\usepackage{cancel}
\usepackage{placeins}
\usepackage{orcidlink}
\usepackage[normalem]{ulem}

\newcommand\bluesout{\bgroup\markoverwith{\textcolor{blue}{\rule[0.5ex]{2pt}{0.4pt}}}\ULon}

\usepackage{hyperref}%[colorlinks=true,citecolor=blue]
\hypersetup{colorlinks=true,citecolor=blue,linkcolor=blue,urlcolor=blue}

\allowdisplaybreaks

\begin{document}
\title{Tomographic Electron Flow in Confined Geometries: \\ 
Beyond the Dual-Relaxation-Time Approximation}

\author{Nitay Ben-Shachar\orcidlink{0009-0002-2989-2295}}
\email{nshachar@caltech.edu}
\affiliation{Department of Physics, California Institute of Technology, Pasadena, California 91125, USA}
\affiliation{School of Mathematics and Statistics, The University of Melbourne, Victoria 3010, Australia}

\author{Johannes Hofmann\orcidlink{0000-0002-0667-2452}}
\email{johannes.hofmann@physics.gu.se}
\affiliation{Department of Physics, Gothenburg University, 41296 Gothenburg, Sweden}
\affiliation{Nordita, Stockholm University and KTH Royal Institute of Technology, 10691 Stockholm, Sweden}

\date{\today}

\begin{abstract}
Hydrodynamiclike electron flows are typically modeled using the Stokes-Ohm equation or a kinetic description that is based on a dual-relaxation-time approximation. Such models assume a short intrinsic mean free path $\ell_e$ due to momentum-conserving electronic scattering and a large extrinsic mean free path $\ell_\text{MR}$ due to momentum-relaxing impurity scattering. This assumption, however, is overly simplistic and falls short at low temperatures, where it is known from exact diagonalization studies of the electronic collision integrals that another large electronic mean free path $\ell_o$ emerges that describes long-lived odd electron modes---this is sometimes known as the tomographic effect. Here, using a matched asymptotic expansion of the underlying Fermi-liquid kinetic equation with multiple electron relaxation times, we derive a transport theory for tomographic flows in arbitrary smooth boundary geometries. Our key results are a set of governing equations for the electron density and electron current, their slip boundary conditions (which are applied at diffuse boundaries), and boundary layer corrections near diffuse edges. We find that the tomographic effect strongly modifies previous hydrodynamic theories for electron flows: In particular, (i) a local equilibrium is established in the bulk, where the flow is governed by Stokes-Ohm-like equations with significant finite-wavelength corrections, (ii) the velocity slip condition at the boundary is strongly modified and not accounted for by the widely used hydrodynamic slip-length condition, (iii) a large kinetic boundary layer of width \mbox{$\sim\sqrt{\ell_e \ell_o}$} arises near diffuse boundaries, which is much larger than in conventional near-hydrodynamic theories, and (iv) all these effects are strongly suppressed by an external magnetic field. We illustrate our findings for electron flow in a channel. The equations derived here provide a comprehensive framework for future studies of tomographic flows in smooth geometries.
\end{abstract}

\maketitle

%\tableofcontents 

\section{Introduction}

In the conventional description of a conductor (the so-called Drude or Ohmic picture of transport), the resistance and other electronic material properties are set by an extrinsic momentum-relaxing mean free path~$\ell_\text{MR}$, which describes the effect of collisions with impurities or phonons. More recently, however, a fundamentally distinct mode of transport has been observed in high-mobility electronic devices marked by a hydrodynamiclike flow of charge carriers~\cite{fritz24,varnavides23}. Here, frequent momentum-conserving electron-electron collisions set the shortest mean free path $\ell_\text{e}$ and are assumed to establish a local thermodynamic equilibrium with a slow spatial transport of the density as well as charge and energy currents~\cite{forster90}. This leads to collective hydrodynamic flow patterns that have now been experimentally resolved, such as current vortices~\cite{aharonsteinberg22} or Poiseuille flow through a channel~\cite{sulpizio19,ku20,vool21,krebs23}. Such hydrodynamic flow depends sensitively on the device geometry.

The state-of-the-art analysis of hydrodynamic electron flows is based on the so-called Stokes-Ohm equation~\cite{fritz24,varnavides23}, which combines the linearized hydrodynamic equations of laminar flow (the Stokes equation) with a direct momentum-relaxing term that accounts for residual impurity scattering. Electron interactions enter in this equation through a term that describes the diffusion of the velocity, with a diffusion constant set by the electron viscosity \mbox{$\eta = v_F \ell_e/4$} (where $v_F$ is the Fermi velocity). This term defines a single characteristic electronic mean free path $\ell_e$ (i.e., a single electronic relaxation timescale \mbox{$\tau_e = \ell_e/v_F$}) as a microscopic parameter. A solution of the Stokes-Ohm equation in finite device geometries with appropriate boundary conditions~\cite{kiselev19,moessner19} can then be used to interpret measurements in the hydrodynamic regime. The corresponding microscopic kinetic picture, from which the Stokes-Ohm equation is derived in the hydrodynamic limit, is based on the so-called dual-relaxation-time approximation, originally introduced in this context by~\textcite{dejong95}. Fundamentally, both frameworks are semiclassical descriptions that treat the flow of thermally excited electron quasiparticles in a device (such as through a wire) in response to an electric field in the same way as the flow of a gas in a conduit (such as through a pipe) in response to a pressure gradient~\cite{knudsen09}. In this way, the dual-relaxation-time kinetic theory can capture a crossover between Ohmic, ballistic, and hydrodynamic transport regimes.

\begin{figure}
    \centering
    \includegraphics{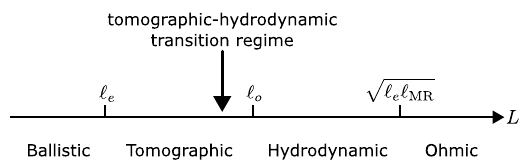}
    \caption{
    Sketch of different flow regimes as a function of the macroscopic device length scale $L$, with an odd-even effect in the quasiparticle mean free paths $\ell_e$ and $\ell_o$ and additional weak disorder, i.e., \mbox{$\ell_e \ll \ell_o \ll \ell_{\rm MR}$}. For small device dimensions, the mean free path of all modes exceeds the system size and the flow is ballistic. With increasing length scale, the electron gas enters the tomographic transport regime, for which the mean free path of even-parity modes is now small compared to the device dimension while odd-parity modes remain ballistic. Fully hydrodynamic flow, where all electronic mean free paths are small compared to the system size, is established only at even larger scales. Finally, a further transition from hydrodynamic to conventional Ohmic transport dominated by impurity scattering occurs at even larger system scales that exceed the Gurzhi length scale $\sqrt{\ell_e \ell_\text{MR}}$. We are primarily interested in describing the tomographic regime and the tomographic-hydrodynamic transition regime.}
    \label{fig:lengthscales}
\end{figure}

While this description works well at temperatures that are a sizable fraction of the Fermi temperature $T_F$ (i.e., approaching the nondegenerate limit), it is expected to require significant modification at low temperatures due to Pauli blocking~\cite{baker24}: Here, microscopically, relaxation proceeds by ``head-on'' electron-electron collisions, but these processes relax only the distribution function's even modes~\cite{laikhtman92,nilsson05}, i.e., deformations of the electron distribution that are symmetric with respect to the electron momentum and described by the parity-symmetric part 
\begin{align}
[h]_e=\frac{1}{2} \Bigl( h(v_i)+h(-v_i) \Bigr) \label{eq:he}
\end{align}
of the electron distribution $h(v_i)$, where $v_i$ is the electron velocity. In contrast, the distribution function's odd modes---i.e., deformations of the Fermi surface that are antisymmetric with respect to the electron momentum, described by
\begin{align}
[h]_o=\frac{1}{2} \Bigl( h(v_i)-h(-v_i) \Bigr) \label{eq:ho}
\end{align}
---are not relaxed by head-on collisions and become anomalously long-lived~\cite{gurzhi95,ledwith19,ledwith17,hofmann23,nilsson24,kryhin23,kryhin25,musser26}. This implies that a third characteristic length scale~$\ell_o$ arises that describes the relaxation of these odd modes and that is typically significantly larger than the microscopic electronic length~$\ell_e$ (see Fig.~\ref{fig:lengthscales}). The suppression of the relaxation of these odd modes has been dubbed the ``tomographic'' or the ``odd-even'' effect. Several recent findings have already been attributed to this effect, including a linear-in-$T$ dependence of the electron viscosity~\citep{zeng24}, an odd-even effect in the linewidth of cyclotron resonances~\cite{moiseenko24}, and low-temperature details of the superballistic conductance compared to hydrodynamic predictions~\citep{estrada25,starkov25}.

%++++++++++++++++++++++++++++++++++++++++
%++++++++++++++++++++++++++++++++++++++++

The presence of long-lived odd modes calls into question the standard hydrodynamic or dual-relaxation-time description of electron flows: In the absence of spontaneous symmetry breaking, hydrodynamics relies on describing the dynamics of the system's conserved quantities, which are charge, momentum, and energy. For tomographic flows, one would expect the long-lived odd modes to be part of such a description and that kinetic effects beyond the hydrodynamic limit become important. Indeed,~\citet{ledwith19} demonstrated that, for bulk flows in the tomographic regime, the current density is coupled to these higher-order odd modes, which gives rise to a wavelength-dependent effective viscosity for near-ballistic flows. However, past studies did not rigorously model scattering at device edges, which was instead accounted for by applying no-slip boundary conditions~\cite{ledwith19,nazaryan24}. While such no-slip boundary conditions are often associated with microscopically rough device edges, which give rise to a diffuse scattering condition on the electron distribution at the device boundaries, a simple link between the microscopic diffuse boundary scattering and a no-slip boundary condition on the electron velocity holds only in a strict hydrodynamic theory and (as we show in this paper) is not true, in general. An unambiguous identification of tomographic transport thus requires a description not only of bulk finite-wavelength corrections, but also of the distinct tomographic phenomena in confined geometries, including modified slip boundary conditions.
 
In this paper, we provide a rigorous asymptotic description of tomographic flow by performing a matched asymptotic expansion of the underlying kinetic equation. The expansion applies to an arbitrary smooth boundary geometry that scatters electrons diffusely, and we further assume a parabolic dispersion relation for simplicity. We limit the analysis to the linear-response regime, which applies whenever flow velocities are small compared to the Fermi velocity (cf.~Appendix~\ref{app:linearization}) and for which the collision integral can be diagonalized using an angular harmonic basis expansion~\citep{nilsson24}. The small parameter in our expansion is the ratio of the mean free path $\ell_e$ of hydrodynamic modes to the macroscopic scale $L$ of the flow (such as the device dimension):
\begin{equation}
    k_e = \frac{\ell_e}{L} \ll 1 , \label{eq:defknudsen}
\end{equation}
where the last inequality holds, by definition, in the \mbox{(near-)}hydrodynamic flow regime of interest here (but does not apply, for example, for ballistic transport). The parameter in Eq.~\eqref{eq:defknudsen} is known as the (even-mode) Knudsen number~\cite{sone02,sone07}. Here, we employ the matched asymptotic expansion technique to obtain an asymptotic theory for tomographic electron flows, which satisfy
\begin{align}
    \ell_e\ll L \lesssim \ell_o \ll \ell_\text{MR} . \label{eq:tomoLengths}
\end{align}
This length scale separation is illustrated graphically in Fig.~\ref{fig:lengthscales}, along the scale $L$ at which the system exhibits ballistic, tomographic, hydrodynamic, and Ohmic flows, respectively. The existence of a near-continuum theory for tomographic flows is not obvious, since the presence of a large ballistic length scale $\ell_o$ might seem to preclude any systematic hydrodynamic expansion. However, as we show, we can describe tomographic flows via an expansion distinct from that used for conventional near-hydrodynamic flows, using $\sqrt{k_e}$ rather than $k_e$ as the expansion parameter. Our analysis captures the leading-order deviation from the strict limit dictated by Eq.~\eqref{eq:tomoLengths} and, thus, extends our theory's applicability to flows where $\ell_e, L, \ell_o$, and $\ell_{\rm MR}$ are not widely separated. This is the regime appropriate to experiments~\citep{moiseenko24,zeng24}, where the relevant microscopic length scales achieved typically do not deviate by more than $1$ or $2$ orders of magnitude. In fact, the measurements reported in Ref.~\citep{moiseenko24} are consistent with an odd-mode mean free path that is no more than $3$ times the even-mode mean free path, $\ell_o/\ell_e\approx 3$.

%++++++++++++++++++++++++++++++++++++++++
\begin{figure}
\centering
\includegraphics{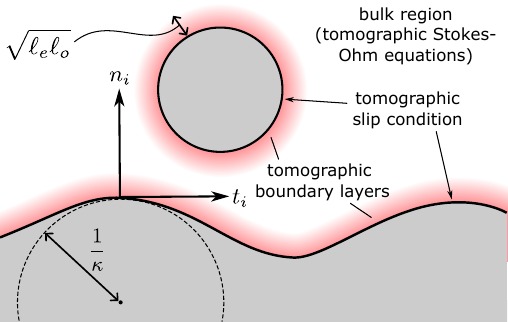}
\caption{An example geometry with the flow domain shown in white and boundaries in gray. The sample edge has a local curvature $\kappa$ and is described by boundary normal and tangent unit vectors $n_i$ and $t_i$, respectively. We point out new effects of tomographic flows compared to a hydrodynamic description: (i) The flow in the bulk region is governed by the tomographic Stokes-Ohm equations, which are derived herein. (ii) These are subject to new tomographic slip conditions which are applied at the boundaries (as indicated). (iii) The approximate extent of the anomalously large tomographic boundary layers, where the bulk description breaks down and requires further tomographic corrections, is indicated by a red shaded region.}
\label{fig:bdryGeo}
\end{figure}
%++++++++++++++++++++++++++++++++++++++++

Our solution of the kinetic equation in this limit reveals four striking phenomena for tomographic flows (see Fig.~\ref{fig:bdryGeo}): First, sufficiently far from the diffuse edges (which we term the bulk region), the flow is described by Stokes-Ohm-like equations with higher-order derivative terms associated with finite-wavelength effects, known as rarefaction phenomena in the hydrodynamics literature~\citep{sone02,sone07}. Second, the boundary conditions for these bulk equations exhibit a strong deviation from the widely used slip-length velocity boundary condition. In particular, the leading-order velocity slip condition for the bulk flow depends on the local curvature of the device edge, the applied magnetic field strength, and second derivatives of the velocity at the boundary. These additional velocity slip contributions arise due to a strong deviation from equilibrium facilitated by the long-lived odd modes, as evidenced by the explicit dependence of the velocity slip condition on both the even- and odd-mode mean free paths. Third, near any diffuse edge, an anomalously large kinetic boundary layer of thickness~\mbox{$\sim\sqrt{\ell_e \ell_o}$} arises (which we term the ``tomographic boundary layer''), where the flow is not described by the above-mentioned bulk expansion. A separate expansion is required to describe the tomographic phenomena within this layer. Last, all of the above-mentioned tomographic phenomena are found to be suppressed by an applied magnetic field, from which the relaxation of the odd modes can be easily extracted.

We remark that the matched asymptotic expansion technique was applied recently to conventional near-hydrodynamic electron flows that do not exhibit the tomographic effect~\cite{benshachar25a,benshachar25b}, i.e., for which \mbox{$\ell_e=\ell_o \ll L \ll \ell_\text{MR}$} as opposed to Eq.~\eqref{eq:tomoLengths}. As expected, such an analysis does not capture the tomographic phenomena discussed above. This is most clearly illustrated by the width of the kinetic layer obtained here (i.e., the tomographic layer), which scales with $\sqrt{\ell_e \ell_o}$ and is, thus, anomalously large compared to the Knudsen layer of Refs.~\cite{benshachar25a,benshachar25b}, whose width scales with $\ell_e$. Indeed, as outlined above, the new scaling of the tomographic layer requires a separate analysis that is fundamentally distinct from the asymptotic expansion of hydrodynamic flows. We note that the analysis here assumes that the tomographic boundary layers do not overlap (i.e., \mbox{$\sqrt{\ell_e \ell_o} < L$}), as indicated in Fig.~\ref{fig:bdryGeo}. When \mbox{$L \lesssim \sqrt{\ell_e \ell_o}$}, i.e., in the tomographic-ballistic regime (cf.~Fig.~\ref{fig:lengthscales}), a different analysis is required to account for strong deviations from local equilibrium throughout the spatial domain, as in Refs.~\citep{ledwith19,rostami24,kryhin25}.

The remainder of the paper is structured as follows: We begin in Sec.~\ref{sec:kinetic_theory} with a description of the underlying Fermi-liquid kinetic theory of electron transport and outline the expansion of the kinetic equation for tomographic flows. Section~\ref{sec:summary} then lists our key findings. These include the governing equations for the mean velocity and electrochemical potential, their boundary conditions, and tomographic boundary layer corrections. These equations provide a comprehensive framework for the study of tomographic flows and constitute the central result of our work. They are valid for flow in arbitrary devices with a smooth edge geometry. Section~\ref{sec:MAE} provides a detailed derivation of the asymptotic solution to the governing kinetic equation, which gives the above-mentioned bulk equations, boundary conditions, and tomographic boundary layer corrections. We illustrate the use of the asymptotic theory by solving for the flow in a channel in Sec.~\ref{sec:channel_flow}, discussed in detail in Ref.~\cite{benshachar25ShortPaper}, which provides extended results on the flow profile and the Fermi surface deformation. The paper concludes with a summary and outlook in Sec.~\ref{eq:conclusion}. Details of some calculations are relegated to five Appendixes.

\section{Deriving transport equations from Fermi-liquid theory}\label{sec:kinetic_theory}

We consider a kinetic equation in the linear-response regime with separate relaxation rates for the distribution function's even and odd modes as well as an additional rate of momentum-relaxing scattering. While microscopic studies of the electron collision integral indicate nearly uniform decay rates for all even-parity modes, odd-parity decay rates depend on the angular harmonic number of the mode~\cite{nilsson24}. In the following analysis, we therefore assume a constant relaxation rate for all even-parity modes and allow for general models of odd-mode relaxation. In dimensionless form, we consider 
\begin{align}
    &v_i \frac{\partial h}{\partial x_i} + 2k_e v_i E_i - \frac{1}{r_c}v_i^\perp \frac{\partial h}{\partial v_i} \nonumber \\[1ex] 
    &= - \frac{1}{k_e} \left(\left[h\right]_e - \delta \mu\right) - \frac{1}{k_o} J_\text{odd}(\left[h\right]_o) - \frac{k_e}{G^2}\left(h-\delta \mu\right) , \label{eq:BTE}
\end{align}
where repeated indices are summed over and parameters are defined below. This is the main equation that we analyze in this paper for arbitrary geometries with diffuse boundary conditions. For a derivation of Eq.~\eqref{eq:BTE} from the electron Boltzmann equation under the assumption of parabolic dispersion, we refer to  Appendix~\ref{app:linearization}. 

The left-hand side of the kinetic equation~\eqref{eq:BTE} is the streaming term, where the spatial and velocity coordinates, $x_i$ and $v_i$, respectively, are scaled by the macroscopic length scale $L$ and the Fermi velocity $v_F$, and the dimensionless electric-field unit vector $E_i$ has been scaled by the magnitude of the applied electric field $\mathcal{E}$. The distribution function $h$ describes the local deformation of the Fermi surface at the point $x_i$, which is parametrized by the orientation $\theta$ of the velocity vector $v_i$. Explicitly,
\begin{align}
	h(v_i,x_i) = \sum_{m=0}^\infty \bigl[a_m(x_i) \cos(m\theta) + b_m(x_i) \sin(m\theta)\bigr] , \label{eq:angularexpansionh}
\end{align}
where even and odd values of $m$ set the even and odd components of the distribution function $\left[h\right]_e$ and $\left[h\right]_o$; cf.~Eqs.~\eqref{eq:he} and~\eqref{eq:ho}. The zeroth and first moments of the distribution function give the macroscopic variables of the flow, which are the scaled nonequilibrium electrochemical potential 
\begin{align}
    \delta \mu = \int_{-\pi}^\pi \frac{d\theta}{2\pi} \, h \label{eq:momentsdensity}
\end{align}
and the mean velocity 
\begin{align}
    u_i = \int_{-\pi}^\pi \frac{d\theta}{2\pi} \, (v_i h). \label{eq:momentsvelocity}
\end{align}
These dimensionless quantities are scaled by \mbox{$E_F U/v_F$} and \mbox{$U=e\mathcal{E}L^2/(v_F \ell_e m^*)$}, respectively, where $e$ is the fundamental charge, $m^*$ is the effective mass of the electrons, and $E_F$ is the Fermi energy. The streaming term contains an advective term, and we also include both external electric and magnetic fields, where we introduce the shorthand \mbox{$v_i^\perp = \varepsilon_{ij} v_j$} and the scaled cyclotron radius
\begin{align}
    r_c = \frac{m^*v_F}{e\mathcal{B}L} = \frac{R_c}{L} ,
\end{align}
where $\cal B$ is the strength of the applied magnetic field and $R_c$ is the dimensional cyclotron radius.

The right-hand side of Eq.~\eqref{eq:BTE} is the key departure from a conventional description of hydrodynamic transport and models collisions beyond the dual-relaxation-time approximation. Here, the first term describes the relaxation by electron-electron collisions of even modes with even-mode Knudsen number \mbox{$k_e \ll 1$} [cf.~Eq.~\eqref{eq:defknudsen}], where the subtraction in the round bracket excludes the density zero mode. The second term in Eq.~\eqref{eq:BTE} describes the electronic relaxation of odd modes, where we separate the characteristic odd-mode Knudsen number
\begin{align}
k_o &= \frac{\ell_o}{L} ,
\end{align}
where~$\ell_o$ is the longest odd-mode mean free path (associated with the~\mbox{$m=3$} deformation in Eq.~\eqref{eq:angularexpansionh}~\cite{nilsson24}) and $J_\text{odd}(\left[h\right]_o)$ is the general odd-mode collision operator. This operator is diagonal in the angular basis~\eqref{eq:angularexpansionh},
\begin{align}
    J_\text{odd}(\left[h\right]_o)& =\sum_{m \, {\rm odd}} \lambda_m \bigl[a_m(x_i) \cos(m\theta) + b_m(x_i) \sin(m\theta)\bigr] ,
    \label{eq:lambda_m}
\end{align}
with general odd eigenvalues $\lambda_m$. Momentum conservation dictates \mbox{$\lambda_1=0$}, and with our convention \mbox{$\lambda_3 = 1$} [since we separate $k_o$ in Eq.~\eqref{eq:BTE}] and~\mbox{$\lambda_m \geq 1$} otherwise. By Eq.~\eqref{eq:tomoLengths}, $k_o$ is of order unity, in contrast to the even-mode Knudsen number $k_e$, which is small. In a minimal model of tomographic transport, the relaxation rates of all odd modes are assumed constant, i.e., \mbox{$\lambda_3=\lambda_5=\cdots=1$}~\cite{hofmann22}. More refined models based on exact diagonalization of the electron-electron collision operator include a characteristic \mbox{$\sim m^{p}$} dependence (often with \mbox{$p=4$}) of the odd-mode relaxation rates for small $m$, which saturates at the relaxation rate of even-mode damping at large $m$~\citep{ledwith19,rostami24}. Such a transition can be captured in a simple model by a Matthiessen-rule ansatz for the $m$th odd-mode mean free path \mbox{$\ell_m=\ell_e +\ell'/m^p$}, which corresponds to
\begin{align}
    \lambda_m = \frac{K + 3^{-p}}{K+m^{-p}}, \qquad 
    K=\frac{\ell_e}{\ell'}\ll 1, \label{eq:lambda_m_model}
\end{align}
where the numerator is chosen to ensure \mbox{$\lambda_3=1$}. Note that the parameter $K$ here is fixed by the odd- and even-mode Knudsen numbers as \mbox{$K=3^{-p}(k_o/k_e-1)^{-1}$}. We use this model in the following analysis, but we emphasize that our theory is applicable to general odd-parity relaxation models. Finally, the third term in Eq.~\eqref{eq:BTE} accounts for momentum-relaxing processes that are parametrized by the Gurzhi number
\begin{align}
G = \frac{\sqrt{\ell_e \ell_\text{MR}}}{L} .
\end{align}
If we restate the length scale separation in Eq.~\eqref{eq:tomoLengths} in terms of the even- and odd-mode Knudsen numbers as well as the Gurzhi number, we find
\begin{align}
    k_e\ll 1 , \qquad G,k_o \sim 1 . \label{eq:order}
\end{align}
As we show in this paper, the parameter $k_e$ can be used to construct a systematic expansion around the hydrodynamic limit.

Equation~\eqref{eq:BTE} must be solved with appropriate boundary conditions on the microscopic distribution function. Diffuse boundary conditions constrain the distribution of states that are reflected at the edges [i.e., which have a velocity component that points into the flow domain, \mbox{$v_in_i>0$}; cf.~Fig.~\ref{fig:bdryGeo}] as
\begin{align}
    \bigl[ h(v_i) \bigr]_S = -\frac{1}{2}\int_{v_in_i<0}d\theta \, (v_in_i) \bigl[ h(v_i)\bigr]_S , \label{eq:diffuseintro}
\end{align}
where the subscript $S$ indicates that the distribution is evaluated at the edge. This boundary condition expresses that the distribution for all reflected velocities \mbox{$v_i n_i>0$} is completely randomized and, hence, isotropic. The constant on the right-hand side of Eq.~\eqref{eq:diffuseintro} ensures that the velocity component~\eqref{eq:momentsvelocity} normal to the edge vanishes, i.e., that there is no penetration through the boundary. Crucially, scattering at device edges prevents the establishment of a local equilibrium at the boundary. This gives rise to a kinetic boundary layer near any edge over which the nonequilibrium distribution function at the boundary relaxes toward the local equilibrium distribution function in the bulk. As we shall show, for tomographic flows this decay arises from the balance of even-mode-induced diffusion and odd-mode collisions, and so the width of the kinetic boundary layer scales with $\sqrt{\ell_e \ell_o}$. Thus, we term this boundary layer the ``tomographic boundary layer.'' As illustrated in Fig.~\ref{fig:bdryGeo}, the flow domain is split into a bulk region and a boundary layer region (red shaded area). In the bulk region (i.e., at distance scales larger than $\sqrt{\ell_e \ell_o}$ from all boundaries), the flow is dominated by collisions that establish a local equilibrium, giving rise to a hydrodynamiclike ``bulk'' flow. As we show, this bulk flow is governed by Stokes-Ohm-like equations with higher-order derivative corrections.

Formally, we find that the ${\it O}(\sqrt{\ell_e \ell_o})$ width of the tomographic boundary layer requires the distribution function and each of its moments to be expanded in powers of the nonanalytic parameter $\sqrt{k_e}$:
\begin{align}
    \Psi = \Psi^{(0)}_B+\sqrt{k_e}\Psi^{(1)}_B+k_e \left(\Psi^{(2)}_B + \Psi^{(2)}_T\right) + \cdots . \label{eq:keExpansion}
\end{align}
Here, \mbox{$\Psi \in \{h, \delta \mu, u_i\}$} denotes any of the dependent variables, and the subscripts ``$B$'' and ``$T$,'' respectively, correspond to the solution in the bulk and the correction in the tomographic boundary layer (see Fig.~\ref{fig:bdryGeo}), the latter of which decays rapidly away from any diffuse edge. Note that the structure of this expansion differs markedly from the conventional Knudsen expansion of hydrodynamics, for which the expansion parameter is $k_e$. The solution for the macroscopic variables proceeds by substituting the expansions in Eq.~\eqref{eq:keExpansion} into Eq.~\eqref{eq:BTE} for both the bulk and tomographic regions and collecting powers of $\sqrt{k_e}$. This solution is summarized in Sec.~\ref{sec:summary}, with details of the derivation reported in Sec.~\ref{sec:MAE}.

\section{Asymptotic theory for tomographic flows}
\label{sec:summary}

In this section, we present the  asymptotic theory for tomographic flows [i.e., flows that exhibit the length scale separation~\eqref{eq:tomoLengths}, or equivalently satisfy Eq.~\eqref{eq:order}]. The resulting equations are summarized in Table~\ref{tbl:BCs_tomographicCorrections}. As discussed, the asymptotic equations for tomographic flows are obtained by substituting Eq.~\eqref{eq:keExpansion} [Table~\ref{tbl:BCs_tomographicCorrections}, Sec.~1] into Eq.~\eqref{eq:BTE} and collecting powers of $\sqrt{k_e}$. The solution of the bulk distribution function at each order of  $\sqrt{k_e}$ requires the macroscopic variables (i.e., $\delta \mu_B^{(n)}$ and $u_{B|i}^{(n)}$ for each $n$) to satisfy coupled governing equations, which are listed up to ${\it O}(k_e)$ in Table~\ref{tbl:BCs_tomographicCorrections}, Sec.~2 and which consist of incompressibility conditions and tomographic Stokes-Ohm-like equations. The solution for the distribution function in the tomographic layer at each order of $\sqrt{k_e}$ then gives (i)~the appropriate boundary conditions for the above-mentioned bulk equations, which are applied at diffuse edges [Table~\ref{tbl:BCs_tomographicCorrections}, Sec.~3], and (ii)~tomographic corrections to the macroscopic variables, which are non-negligible within the tomographic boundary layer [Table~\ref{tbl:BCs_tomographicCorrections}, Sec.~4]. For a particular device geometry, the asymptotic solutions for $\delta \mu$ and $u_i$ are then obtained by first solving the bulk governing equations listed in Sec.~2 in Table~\ref{tbl:BCs_tomographicCorrections} (subject to their velocity boundary conditions in Sec.~3 in Table~\ref{tbl:BCs_tomographicCorrections}) and then adding the tomographic-layer corrections listed in Sec.~4 in Table~\ref{tbl:BCs_tomographicCorrections} to this bulk solution. We illustrate this procedure in Sec.~\ref{sec:channel_flow} for electron flow in a channel. 

The bulk governing equations, their boundary conditions, and tomographic boundary layer corrections listed in Table~\ref{tbl:BCs_tomographicCorrections} are the main result of this study. They form a complete asymptotic theory of tomographic electron flow up to ${\it O}(k_e)$, which captures phenomena beyond the hydrodynamic limit and which may be utilized to calculate the flow in any device with smooth boundary geometry up to linear order for \mbox{$k_e \ll 1$} (including devices with curved boundaries and with an applied perpendicular magnetic field). Thus, they provide an alternative to computationally expensive numerical solutions of the kinetic equation for electron flows at small~$k_e$, and they also provide crucial analytical insight into the flow structure. Moreover, the microscopic distribution function is expressed in terms of the macroscopic variables in Table~\ref{tbl:hSummary}, and, thus, a solution for the macroscopic variables allows one to reconstruct the full distribution function.

%++++++++++++++++++++++++++++++++++++++++
\begin{table*}[h!]
    \centering
    \begin{tabular}{m{\linewidth}}
    \hline \hline \\ 
%==========================
    1.\hspace{1em} Asymptotic expansion \\
    \hline \\
    \begin{equation} 
    \begin{array}{rl}
        &\displaystyle\delta \mu = \delta \mu_B^{(0)} + \sqrt{k_e} \, \delta \mu_B^{(1)} + k_e \left(\delta \mu_B^{(2)} + \delta \mu_T^{(2)}\right)+\cdots \\[1ex]
        &\displaystyle u_i = u_{B|i}^{(0)} + \sqrt{k_e} \, u_{B|i}^{(1)} + k_e \left(u_{B|i}^{(2)} + u_{T|i}^{(2)}\right)+\cdots
    \end{array}
    \end{equation} \\
%==========================
    2.\hspace{1em} Bulk equations \\\hline \\
    {\begin{subequations}\vspace{0.2cm}
    \label{eq:bulkEqs}
    Incompressibility condition:\vspace{-1cm}
    \begin{align}
        \label{eq:bulkEqn_a}
        \displaystyle \frac{\partial u_{B|i}^{(n)}}{\partial x_i} = 0, \quad n \geq 0 .
    \end{align}
    \noindent Tomographic Stokes-Ohm equations:
    \vspace{-0.2cm}
    \begin{align}
        \label{eq:bulkEqn_b} 
        \displaystyle -\frac{1}{2}\frac{\partial \delta \mu^{(n)}_B}{\partial x_i} = & \displaystyle 
        \frac{1}{r_c} \varepsilon_{ij}u_{B|j}^{(n)} ,\hspace{2.2em} n\in\{0,1\}, \\ 
        \label{eq:bulkEqn_c} 
        \displaystyle - \frac{1}{2}\frac{\partial \delta\mu_B^{(2)}}{\partial x_i}  = & \displaystyle\frac{u_{B|i}^{(0)}}{G^2}-\frac{1}{4}\frac{\partial^2 u_{B|i}^{(0)}}{\partial x_j^2} 
        +\frac{1}{r_c} \varepsilon_{ij}u_{B|j}^{(2)} +E_i , \\
        \label{eq:bulkEqn_d} 
        \displaystyle - \frac{1}{2}\frac{\partial \delta\mu_B^{(3)}}{\partial x_i}  = & \displaystyle\frac{u_{B|i}^{(1)}}{G^2}-\frac{1}{4}\frac{\partial^2 u_{B|i}^{(1)}}{\partial x_j^2} 
        +\frac{1}{r_c} \varepsilon_{ij}u_{B|j}^{(3)} , \\
        \label{eq:bulkEqn_n2}
        \displaystyle - \frac{1}{2}\frac{\partial \delta\mu_B^{(4)}}{\partial x_i}  = & \displaystyle \frac{u_{B|i}^{(2)}}{G^2}-\frac{1}{4}\frac{\partial^2 u_{B|i}^{(2)}}{\partial x_j^2} 
        +\frac{1}{r_c} \varepsilon_{ij}u_{B|j}^{(4)} + \frac{1}{2r_c}\varepsilon_{ij}\frac{\partial^2 u_{B|j}^{(0)}}{\partial x_k^2}- \frac{k_o}{16(1+(3k_o/r_c)^2)}\biggl(\frac{\partial^4 u_{B|i}^{(0)}}{\partial x_k^2\partial x_l^2} -\frac{3k_o}{r_c}\varepsilon_{ij} \frac{\partial^4 u_{B|j}^{(0)}}{\partial x_k^2\partial x_l^2} \biggr).
    \end{align}
    \end{subequations}} \\
%==========================
    3.\hspace{1em} Boundary conditions \\
    \hline \\
    \vspace{0.2cm}
    {\begin{subequations}
    \label{eq:bulkBCs}
    No penetration:\vspace{-0.7cm}
    \begin{align} 
        \Bigl[u_{B|i}^{(0)}n_i \Bigr]_S &= 0, \qquad 
        \Bigl[u_{B|i}^{(1)}n_i \Bigr]_S = 0, \qquad 
        \Bigl[u_{B|i}^{(2)}n_i \Bigr]_S = 0 . \label{eq:nopenetrationbc} \\ \nonumber
    \end{align}
    No slip and tomographic slip: \vspace{-0.75cm}
    \begin{align} 
        \label{eq:bulkBCns}
        \Bigl[ u_{B|i}^{(0)}t_i \Bigr]_S = 0, \qquad &\Bigl[u_{B|i}^{(1)}t_i \Bigr]_S = 0, \\
        \label{eq:bulkBCslip}
        \displaystyle \Bigl[u_{B|i}^{(2)}t_i \Bigr]_S = \displaystyle  
        - \frac{32}{15\pi} \Bigl[n_it_j S_{B|ij}^{(0)} \Bigr]_S
        - \frac{k_o}{2(1+(3k_o/r_c)^2)} \Biggl(\kappa \Bigl[n_it_j &S_{B|ij}^{(0)} \Bigr]_S
        - \frac{1}{2}\biggl[t_in_jn_k \frac{\partial S_{B|ij}^{(0)}}{\partial x_k} \biggr]_S + \frac{9}{4}\frac{k_o}{r_c} \biggl[n_in_jn_k\frac{\partial S_{B|ij}^{(0)}}{\partial x_k} \biggr]_S \Biggr).
    \end{align}
    \end{subequations}} \\
%==========================
    4.\hspace{1em} Tomographic boundary layer \\
    \hline \\
    {\begin{subequations}
    Rescaled tomographic boundary layer normal  coordinate:\vspace{-0.75cm}
    \begin{align}
        \hspace{2cm}\chi = \frac{x_in_i}{\sqrt{k_ek_o}} .\label{eq:defchi}
    \end{align}%
    Tomographic boundary layer corrections:
    \label{eq:tomoCorrections}
    \begin{align}
        \label{eq:nopenetrationT} 
        u_{T|i}^{(2)}n_i &= 0, \\ 
        \label{eq:tomoLayerCorrection_u}
        \displaystyle u_{T|i}^{(2)}t_i 
        &= \displaystyle \mathcal{Y}_0(\chi;\tfrac{k_o}{r_c}) \Bigl[n_it_j S_{B|ij}^{(0)} \Bigr]_S 
        + k_o\mathcal{Y}_1(\chi;\tfrac{k_o}{r_c})\kappa \Bigl[n_it_j S_{B|ij}^{(0)} \Bigr]_S
        - \frac{k_o}{2} \mathcal{Y}_1(\chi;\tfrac{k_o}{r_c}) \biggl[t_in_jn_k \frac{\partial S_{B|ij}^{(0)}}{\partial x_k} \biggr]_S
        + k_o \mathcal{Y}_2(\chi;\tfrac{k_o}{r_c}) \biggl[n_in_jn_k\frac{\partial S_{B|ij}^{(0)}}{\partial x_k} \biggr]_S , \\[2ex]
        \label{eq:tomoLayerCorrection_mu2} 
        \delta \mu_T^{(2)} &= 0 , \\
        \label{eq:tomoLayerCorrection_mu}
        \displaystyle \frac{\partial \delta \mu_T^{(3)}}{\partial \chi} 
        &= \displaystyle  \frac{2\sqrt{k_o}}{r_c} u_{T|i}^{(2)} t_i + \frac{1}{\sqrt{k_o}}\mathcal{T}_0(\chi;\tfrac{k_o}{r_c}) \Bigl[n_it_j S_{B|ij}^{(0)} \Bigr]_S 
        + \sqrt{k_o}\mathcal{T}_1(\chi;\tfrac{k_o}{r_c})\kappa \Bigl[n_it_j S_{B|ij}^{(0)} \Bigr]_S
        - \frac{\sqrt{k_o}}{2} \mathcal{T}_1(\chi;\tfrac{k_o}{r_c}) \biggl[t_in_jn_k \frac{\partial S_{B|ij}^{(0)}} {\partial x_k}\biggr]_S\nonumber \\
        &\quad\displaystyle + \sqrt{k_o} \mathcal{T}_2(\chi;\tfrac{k_o}{r_c}) \biggl[n_in_jn_k\frac{\partial S_{B|ij}^{(0)}}{\partial x_k} \biggr]_S .  
        \end{align}
    \end{subequations}}\\[0ex]
    \hline
    \hline
    \end{tabular}
    \caption{Summary of the  tomographic flow equations: 
    1. The electrochemical potential and the flow velocity are expanded in powers of $\sqrt{k_e}$, where \mbox{$k_e\ll 1$} is the small Knudsen number. The expansion consists of a bulk contribution (subscript ``$B$'') and, starting at order ${\it O}(k_e)$, a tomographic boundary layer contribution (subscript ``$T$''). 
    2. The bulk macroscopic variables obey incompressibility and  ``tomographic Stokes-Ohm'' equations. 
    3. Boundary conditions for the bulk flow equations listed in 2 for diffuse microscopic boundary scattering. \mbox{$S_{B|ij}^{(n)}$} is the rate-of-strain tensor in the bulk region [Eq.~\eqref{eq:stresstensor}], and $\kappa$ is the local curvature of the boundary. All bulk quantities are evaluated at the boundary.
    4. Tomographic boundary layer corrections to the velocity, \mbox{$u_{T|i}^{(n)}$}, and Hall field, \mbox{$\partial \delta \mu_T^{(n)}/\partial \chi$}, up to order \mbox{${\it O}(k_e)$}. The tomographic boundary-layer functions $\mathcal{Y}_0$, $\mathcal{Y}_1$, $\mathcal{Y}_2$, $\mathcal{T}_0$, $\mathcal{T}_1$, and $\mathcal{T}_2$ are independent of the geometry and are evaluated in Sec.~\ref{sec:truncatedMomentExp} (see also Figs.~\ref{fig:tomoFun} and~\ref{fig:tomoFun2}).}
    \label{tbl:BCs_tomographicCorrections}
 \end{table*}
%++++++++++++++++++++++++++++++++++++++++

%++++++++++++++++++++++++++++++++++++++++
\begin{table}[t]
    \centering
    \begin{tabular}{m{\linewidth}}
    \hline \hline \\
%==========================
    1.\hspace{1em} Asymptotic expansion \\
    \hline
    \begin{equation} 
        \label{eq:expansionhB}
        \displaystyle h = h_B^{(0)} + \sqrt{k_e} h_B^{(1)} + k_e \left(h_B^{(2)} + h_T^{(2)}\right)+\cdots
    \end{equation} \\
%==========================
    2.\hspace{1em} Bulk \\
    \hline 
    {\begin{subequations}
    \label{eq:hBulk}
    \begin{align}
        \label{eq:hBulk_n01}
        h^{(n)}_B &= \delta \mu^{(n)}_B + 2v_i u_{B|i}^{(n)} , \quad n\in\{0,1\}, \\
        \label{eq:hBulk_n23}
        h^{(n)}_B &= \delta \mu^{(n)}_B + 2v_i u_{B|i}^{(n)} - 2v_iv_j \frac{\partial u_{B|i}^{(n-2)}}{\partial x_j} \nonumber \\
        &\hspace{-1em} + \frac{\partial^2 u_{B|i}^{(n-2)}}{\partial x_j\partial x_k} \Bigg[\frac{2k_o}{1+(3k_o/r_c)^2} \left(v_iv_jv_k-\frac{1}{4} v_i \delta_{jk}\right) \nonumber \\
        &\hspace{-1em}-\frac{6k_o^2/r_c}{1+(3k_o /r_c)^2}\left(v_i^\perp v_j^\perp v_k^\perp - \frac{1}{4}v_i^\perp \delta_{jk}\right)\Bigg] , \quad n\in\{2,3\} .
    \end{align}
    \end{subequations}}
%==========================
    3.\hspace{1em} Tomographic boundary layer correction \\
    \hline
    {\begin{align}
        \label{eq:hT_form}
        h_T^{(2)} &= f_0(\chi,\theta;\tfrac{k_o}{r_c}) \Bigl[ n_it_j S_{B|ij}^{(0)} \Bigr]_S \nonumber \\[0.5ex]
        &\quad + k_o \kappa f_1(\chi,\theta;\tfrac{k_o}{r_c}) \Bigl[ n_it_j S_{B|ij}^{(0)} \Bigr]_S \nonumber \\
        &\quad -\frac{k_o}{2} f_1(\chi,\theta;\tfrac{k_o}{r_c}) \biggl[ t_in_jn_k \frac{\partial S_{B|ij}^{(0)}}{\partial x_k} \biggr]_S \nonumber \\
        &\quad + k_o f_2(\chi,\theta;\tfrac{k_o}{r_c}) \biggl[ n_in_jn_k \frac{\partial S_{B|ij}^{(0)}}{\partial x_k} \biggr]_S .
    \end{align}}
    Tomographic equation:\vspace{-0.45cm}\\
    {\begin{align}
        \label{eq:GEtomo2}
        &(v_in_i)^2 \frac{\partial^2 h_T^{(2)}}{\partial \chi^2} +\frac{k_o}{r_c} v_k^\perp \frac{\partial h_T^{(2)}}{\partial v_k} \nonumber \\[1ex]
        &\quad = J_\text{odd}(h_T^{(2)}) + \sqrt{k_o} v_in_i \frac{\partial \delta \mu_T^{(3)}}{\partial \chi}.
    \end{align}}\\[-2ex]
    \hline 
    \hline 
    \end{tabular}
    \caption{Microscopic distribution function~$h_B^{(n)}$ at each order of the asymptotic expansion and its ${\it O}(k_e)$ correction in the tomographic boundary layer, $h_T^{(2)}$. The functions $f_0$, $f_1$, and $f_2$ are independent of the geometry and solved numerically in Sec.~\ref{sec:truncatedMomentExp}.
    }
    \label{tbl:hSummary}
\end{table}
%++++++++++++++++++++++++++++++++++++++++

In Table~\ref{tbl:BCs_tomographicCorrections}, the geometry of the boundary, which is assumed arbitrary and smooth in this study, is described by a unit normal vector $n_i$ that points into the flow domain and a unit tangent vector $t_i$. We orient the tangent vector by rotating the normal vector by $90^\circ$ clockwise as seen from the positive $z$ direction, i.e.,~\mbox{${\bf t} = {\bf n} \times \hat{\bf z}$} (see Fig.~\ref{fig:bdryGeo}). These satisfy the general geometric relations~\citep{nassios12}
\begin{subequations}
\begin{align}
	n_j \frac{\partial n_i}{\partial x_j} &= 0, & n_j \frac{\partial t_i}{\partial x_j} &= 0, \\
	t_j \frac{\partial n_i}{\partial x_j} &= \kappa t_i, & t_j \frac{\partial t_i}{\partial x_j} &= - \kappa n_i ,
\end{align} \label{eq:kappaDefn}%
\end{subequations}
where $\kappa$ is the local (scaled) curvature of the edge, which is defined to be negative when the center of curvature lies in the electron flow domain. In addition, we define the rate-of-strain tensor at each order
\begin{align}
    S_{B|ij}^{(n)}= -\biggl(\frac{\partial u_{B|i}^{(n)}}{\partial x_j} + \frac{\partial u_{B|j}^{(n)}}{\partial x_i}\biggr) . \label{eq:stresstensor}
\end{align}

We next describe key properties of the asymptotic theory, which are embodied in the bulk governing equations [Sec.~\ref{sec:bulkgoverningequations}], their boundary conditions [Sec.~\ref{sec:boundaryconditions}], the tomographic boundary layer corrections [Sec.~\ref{sec:tomographicboundary}], and the reconstructed electron distribution function [Sec.~\ref{sec:fulldistribution}].

\subsection{Bulk governing equations}\label{sec:bulkgoverningequations}

We begin by describing the bulk governing equations for the electrochemical potential and the velocity profile, which are summarized in Eqs.~\eqref{eq:bulkEqn_a}--\eqref{eq:bulkEqn_n2} in Table~\ref{tbl:BCs_tomographicCorrections}. The first set of constraint equations [Eq.~\eqref{eq:bulkEqn_a}] enforces a divergence-free bulk velocity at all orders in $k_e$, corresponding to incompressible electron flow in spite of generalized relaxation times. The second set of constraint equations [Eqs.~\eqref{eq:bulkEqn_b}--\eqref{eq:bulkEqn_n2}] describes macroscopic momentum transport and includes both the bulk Hall field $-\partial \delta \mu_B/\partial x_i$ and the velocity profile. While in conventional hydrodynamics these latter equations are the standard Stokes-Ohm equations, we find here that such a description is no longer valid, in general, and is modified by higher-derivative corrections.

First, we obtain the widely used Stokes-Ohm equations [Eqs.~\eqref{eq:bulkEqn_b}--\eqref{eq:bulkEqn_d}] at leading order in the Knudsen number \mbox{$k_e\ll 1$}. These are characterized by a balance of drag from residual momentum-relaxing collisions, shear viscous forces, the magnetic force, and the electric force. From the second term on the right-hand side in Eq.~\eqref{eq:bulkEqn_c}, we read off the (dimensional) shear viscosity of the electron fluid  $v_F\ell_e/4$~\citep{alekseev16,scaffidi2017}. A Hall viscosity term does not arise at this order in the Knudsen number expansion. The appearance of the Stokes-Ohm equation is a consequence of a local equilibrium being established to leading order in the bulk; i.e., microscopically, the Fermi surface deformation contains only a local change in the density and the center of mass.

Second, at order \mbox{${\it O}(k_e)$}, the Hall viscosity arises in the bulk equations [the second-to-last term \mbox{$\sim \varepsilon_{ij}\partial^2 u_{B|j}^{(0)}/\partial x_k^2$} in Eq.~\eqref{eq:bulkEqn_n2}]. This corresponds to the well-known dimensional Hall viscosity of $v_F\ell_e^2/(2R_c)$~\citep{alekseev16,scaffidi2017}. However, at the same order, previously unreported finite-wavelength corrections (i.e., rarefaction phenomena) are also found in the bulk equations in the form of higher-derivative terms \mbox{$\sim \partial^4/\partial x_k^2\partial x_l^2$} [the last term in round brackets in Eq.~\eqref{eq:bulkEqn_n2}]. These corrections are linear in the odd-mode Knudsen number $k_o$ and are, thus, directly associated with the long-lived odd modes of the distribution function. Microscopically, the Fermi surface deformation at this order involves even-parity shear deformations (\mbox{$m=2$}) and the longest-lived odd mode (\mbox{$m=3$}); hence, these results are independent of a particular model of odd-mode scattering.

\subsection{Boundary conditions on the bulk solution}\label{sec:boundaryconditions}

The bulk governing equations~\eqref{eq:bulkEqs} for $\delta \mu_B^{(n)}$ and $u_{B|i}^{(n)}$ must be solved with appropriate boundary conditions, which follow from the microscopic diffuse boundary condition~\eqref{eq:diffuseintro} on the distribution function. The velocity-slip boundary conditions for the bulk governing equations obtained in this way are summarized in Eqs.~\eqref{eq:nopenetrationbc}--\eqref{eq:bulkBCslip} in Table~\ref{tbl:BCs_tomographicCorrections}. As discussed above, the diffuse boundary scattering~\eqref{eq:diffuseintro} imposes a no-penetration boundary condition [Eq.~\eqref{eq:nopenetrationbc}] on the normal component of the velocity up to ${\it O}(k_e)$, which prevents flow through the boundaries. The remainder of the discussion, therefore, focuses on the boundary condition for the tangential component of the velocity, which gives rise to velocity slip.

First, at leading orders ${\it O}(1)$ and ${\it O}(\sqrt{k_e})$ (i.e., for $u_{B|i}^{(0)}$ and $u_{B|i}^{(1)}$), we obtain no-slip boundary conditions [Eq.~\eqref{eq:bulkBCns}], which are normally associated with diffuse reflection in the hydrodynamic limit~\citep{varnavides23}. This is attributed to collisions establishing a local equilibrium throughout the spatial domain at these orders in $\sqrt{k_e}$. Since diffuse reflection forces the reflected electrons into equilibrium with the edge, the hydrodynamic equations require a no-slip condition.

%++++++++++++++++++++++++++++++++++++++++
\begin{figure}
    \centering
    \includegraphics{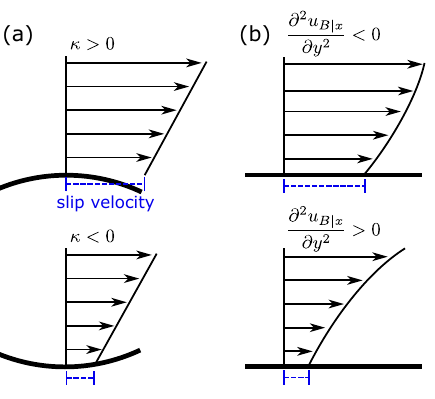}
    \caption{
    Schematic showing the tomographic effect of (a) the boundary curvature and (b) the bulk flow profile (solid black arrows) on the slip-velocity boundary condition (blue dashed line), as dictated by Eq.~\eqref{eq:bulkBCslip}. We indicate boundaries by thick black lines and show bulk flows with equal shear rate (i.e., equal slope at the boundary). In all panels, we draw only the bulk flow, while the full asymptotic velocity profile, which includes the tomographic-layer corrections, is not sketched. 
    }
    \label{fig:slip_schematic}
\end{figure}
%++++++++++++++++++++++++++++++++++++++++

Second, by contrast, at ${\it O}(k_e)$ (i.e., for $u_B^{(2)}$), the velocity boundary conditions deviate from the no-slip condition [Eq.~\eqref{eq:bulkBCslip}]. In particular, a slip-velocity condition arises for the bulk equations (i.e., a condition on the bulk velocity evaluated at the boundary), which is induced by (i) the shear of the leading-order flow [first term in Eq.~\eqref{eq:bulkBCslip}], (ii) the local curvature of the boundary [second term in Eq.~\eqref{eq:bulkBCslip}], (iii) the boundary-normal gradient of the shear rate [third term in Eq.~\eqref{eq:bulkBCslip}], and (iv) the boundary-normal gradient of the normal strain rate [fourth term in Eq.~\eqref{eq:bulkBCslip}], where the rate-of-strain tensor is defined in Eq.~\eqref{eq:stresstensor}. The first term in Eq.~\eqref{eq:bulkBCslip} gives a ``slip-length'' condition, which is often used in the literature~\cite{afanasiev22,palm24,pellegrino16,varnavides23,raichev22}; however, the coefficient \mbox{$32/(15\pi)\approx 0.679$} of this term is approximately 7\% larger than the slip coefficient found for conventional hydrodynamic flows, which is  \mbox{$0.6366\ldots$}~\citep{benshachar25a,benshachar25b,raichev22}. The latter three contributions [round brackets in Eq.~\eqref{eq:bulkBCslip}] are directly associated with the tomographic effect; they do not arise in the leading-order slip boundary conditions of conventional hydrodynamic flows~\citep{benshachar25a,benshachar25b,raichev22}. The last contribution arises in the presence of an applied magnetic field only. The new slip-condition terms arise due to the presence of the long-lived \mbox{$m=3$} mode in the bulk distribution function [the terms in square brackets in Eq.~\eqref{eq:hBulk_n23}, discussed further in Sec.~\ref{sec:fulldistribution}] and, thus, explicitly depend on the odd-mode Knudsen number $k_o$. The appearance of derivatives of the strain in the boundary condition also immediately implies a dependence on the odd-mode mean free path $\ell_o$ on dimensional grounds. Schematics of the slip condition~\eqref{eq:bulkBCslip} are shown in Fig.~\ref{fig:slip_schematic}. In particular, Fig.~\ref{fig:slip_schematic}(a) illustrates the effects of the boundary curvature on the slip [second term in Eq.~\eqref{eq:bulkBCslip}], and Fig.~\ref{fig:slip_schematic}(b) shows the effects of the curvature of the bulk flow velocity at the boundary on the slip [third term in Eq.~\eqref{eq:bulkBCslip}]. Compared with the widely used fixed slip-length condition, the second contribution enhances the slip for convex ($\kappa>0$) boundaries and suppresses it for concave ($\kappa<0$) boundaries. Furthermore, the tomographic contributions to the slip condition in Eq.~\eqref{eq:bulkBCslip} are found to be rapidly suppressed by a magnetic field and decrease in magnitude as a Lorentzian with argument \mbox{$3\ell_o/R_c$}. Together, these phenomena demonstrate that the use of the fixed slip-length boundary condition, which neglects all these corrections, is inadequate for modeling  hydrodynamiclike flows that exhibit the tomographic effect.

%++++++++++++++++++++++++++++++++++++++++
\begin{figure}[b]
    \centering
    \includegraphics{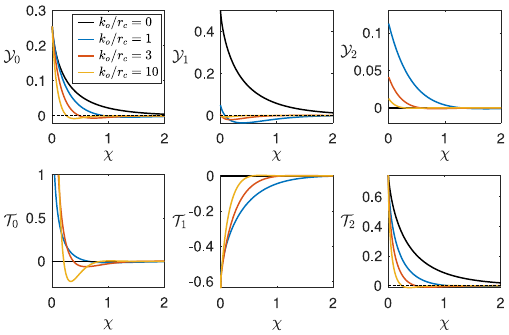}
    \caption{Tomographic boundary-layer functions for the velocity, $\mathcal{Y}_0$, $\mathcal{Y}_1$, and $\mathcal{Y}_2$, and for the Hall field, $\mathcal{T}_0$, $\mathcal{T}_1$, and $\mathcal{T}_2$, as a function of the rescaled spatial coordinate \mbox{$\chi=x_in_i/\sqrt{k_ek_o}$}, for different magnetic fields \mbox{$k_o/r_c\in\{0,1,3,10\}$}. The dashed line marks $0$. These solutions are for a constant odd-mode relaxation rate; solutions for more general odd-mode relaxation models are presented in Appendix~\ref{app:p4_tomoFun}.
    }
    \label{fig:tomoFun}
\end{figure}
%++++++++++++++++++++++++++++++++++++++++

%++++++++++++++++++++++++++++++++++++++++
\begin{figure}
    \centering
    \includegraphics{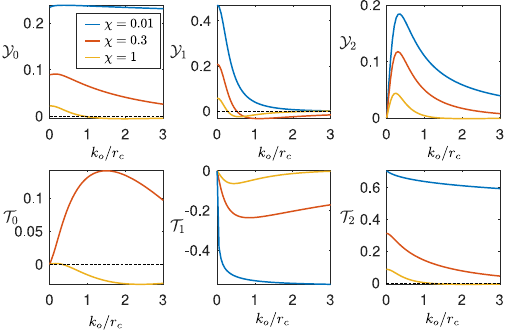}
    \caption{The same as Fig.~\ref{fig:tomoFun} but as a function of magnetic field \mbox{$k_o/r_c$}, for fixed positions \mbox{$\chi\in\{0.01,0.3,1\}$}. The dashed line marks $0$. The function $\mathcal{T}_0$ diverges at \mbox{$\chi=0$} and, thus, is not plotted for \mbox{$\chi=0.01$}. 
    }
    \label{fig:tomoFun2}
\end{figure}
%++++++++++++++++++++++++++++++++++++++++

\subsection{Tomographic boundary layer corrections}\label{sec:tomographicboundary}

The tomographic boundary layer extends over the (dimensional) length scale $\sqrt{\ell_e \ell_o}$. It therefore covers a significantly larger proportion of the flow domain compared with the Knudsen boundary layer that arises in classical hydrodynamic electron flows, which extends over a much shorter length scale $\ell_e$. As we show, the leading-order distribution function in the tomographic boundary layer satisfies Eq.~\eqref{eq:GEtomo2}, which is characterized by a balance of even-mode-induced diffusion, odd-mode collisions, and the magnetic force. Equation~\eqref{eq:GEtomo2} has the same structure as Eq.~(10) in Ref.~\citep{ledwith19}, which motivates the name ``tomographic boundary layer''. Tomographic boundary layer corrections are found at ${\it O}(k_e)$ for the distribution function, velocity profile, and Hall profile (indicated with a subscript $T$), which must be added to the bulk solution. The tomographic boundary layer corrections for the velocity and Hall profiles are listed in Eq.~\eqref{eq:tomoCorrections} in Table~\ref{tbl:BCs_tomographicCorrections}, while the form of the tomographic boundary layer correction for the distribution function is listed in Table~\ref{tbl:hSummary}. Their structure is similar to that of the velocity-slip condition for the bulk solution [Eq.~\eqref{eq:bulkBCslip}]. The functions $\mathcal{Y}_0$, $\mathcal{Y}_1$, $\mathcal{Y}_2$, $\mathcal{T}_0$, $\mathcal{T}_1$, and~$\mathcal{T}_2$, which we term tomographic boundary-layer functions, depend on the magnetic field through the ratio of the odd-mode mean free path to the cyclotron radius and also on the relaxation rates of all odd modes, i.e., \mbox{$\{\lambda_5, \lambda_7, \ldots\}$}. Solutions of these functions (obtained using the numerical algorithm described in Sec.~\ref{sec:truncatedMomentExp}) for $\lambda_m$ given by Eq.~\eqref{eq:lambda_m_model} with \mbox{$p=0$} are shown in Figs.~\ref{fig:tomoFun} and~\ref{fig:tomoFun2}. Analogous solutions for \mbox{$p=4$} in Appendix~\ref{app:p4_tomoFun} show a modest dependence on the scattering rate of higher-order modes.

All tomographic boundary-layer functions become confined near the boundary when the magnetic field strength is increased, and for \mbox{$R_c<\ell_o$}, the boundary layer extends over the (dimensional) length scale \mbox{$\sqrt{\ell_e R_c} < \sqrt{\ell_e \ell_o}$}. Moreover, the tomographic boundary-layer functions $\mathcal{T}_0$, $\mathcal{T}_1$, $\mathcal{Y}_1$, and $\mathcal{Y}_2$ exhibit nonmonotonic behavior with respect to the magnetic field strength (see Fig.~\ref{fig:tomoFun2}).

\subsection{Full electron distribution function}\label{sec:fulldistribution}

Once the macroscopic variables, i.e., the velocity $u_i$ and the Hall field \mbox{$-\partial\delta \mu/\partial x_i$}, are known, it is straightforward to reconstruct the full Fermi surface deformation function $h$. This is summarized in Table~\ref{tbl:hSummary} up to linear order ${\it O}(k_e)$ in the Knudsen number, where the deformation is expanded in powers of $\sqrt{k_e}$ as shown in Eq.~\eqref{eq:keExpansion} or~\eqref{eq:expansionhB} in Table~\ref{tbl:hSummary}. We discuss the structure of these equations next.

The leading-order deformations at ${\it O}(1)$ and ${\it O}(\sqrt{k_e})$ are set by the bulk solution at these orders, Eq.~\eqref{eq:hBulk_n01} in Table~\ref{tbl:hSummary}. They consist of an isotropic change in the electrochemical potential that describes a local change in the electron density as well as a current contribution that displaces the center of mass of the Fermi surface. These are exactly the microscopic deformations that describe hydrodynamic degrees of freedom.

This simple structure of the deformation is modified at ${\it O}(k_e)$ and subsequent orders: The bulk contribution [Eq.~\eqref{eq:hBulk_n23} in Table~\ref{tbl:hSummary}] now contains an additional even-parity deformation that is proportional to the rate-of-strain tensor [the third term in Eq.~\eqref{eq:hBulk_n23}] as well as an odd-parity deformation that is proportional to the curvature of the velocity profile [the last term in square brackets in Eq.~\eqref{eq:hBulk_n23}]. Microscopically, the even-parity term contributes to a Fermi surface deformation with an angular-mode dependence proportional to $\cos(2\theta)$ or $\sin(2\theta)$; i.e., the Fermi surface takes a quadrupolar form that is squeezed along one axis in momentum space but elongated in another. This is a microscopic shear deformation, which is relaxed by even-mode collisions~\cite{gran23}. By contrast, the odd-parity deformations are of the trigonal form $\cos(3\theta)$ or $\sin(3\theta)$. Restoring powers of the Knudsen number, one finds that they are of order ${\it O}(k_ek_o)$ and, thus, induced by repeated odd- and even-mode collisions.

It is not correct, however, to assume that the harmonic expansion truncates at the third harmonic order. Higher deformations enter in the tomographic boundary layer, which also contributes at ${\it O}(k_e)$ [Eq.~\eqref{eq:hT_form} in Table~\ref{tbl:hSummary}]. These corrections depend on the tomographic layer functions $f_0, f_1$, and $f_2$, which solve the tomographic equation~\eqref{eq:GEtomo2}. They are evaluated explicitly in Sec.~\ref{sec:truncatedMomentExp} using a truncated moment expansion in terms of coefficients that describe the projection onto the angular harmonic modes of the deformation. This analysis shows that diffuse boundary scattering induces a Fermi surface deformation that strongly deviates from the hydrodynamic form and involves a large number of higher angular modes. 
The tomographic boundary layer is, thus, sensitive to the odd-mode collision model $J_{\rm odd}$.

\section{Matched asymptotic expansion}\label{sec:MAE}

In this section, we present in detail the matched asymptotic analysis of the kinetic equation~\eqref{eq:BTE} order by order in \mbox{$\sqrt{k_e}$} and show how to derive the results in Tables~\ref{tbl:BCs_tomographicCorrections} and~\ref{tbl:hSummary}. In the bulk region, the distribution function and each of its moments are expanded in a regular perturbation expansion in $\sqrt{k_e}$ with an isotropic length scale (that is, all spatial coordinates are scaled by $L$), as per Eq.~\eqref{eq:keExpansion}. However, as discussed, near diffuse scattering boundaries, the distribution function varies appreciably over the (dimensional) length scale $\sqrt{\ell_e \ell_o}$, which is not captured by the bulk expansion. This is also apparent from the inability of the bulk distribution function to satisfy the diffuse reflection condition at ${\it O}(k_e)$. A tomographic boundary layer correction, which is appreciable within this length scale, must be added for diffuse reflection to be satisfied (see Fig.~\ref{fig:bdryGeo}). Here, the spatial coordinate normal to the boundary is scaled by $\sqrt{\ell_e \ell_o}$, while the coordinate tangential to the boundary is scaled by $L$. We, thus, define the rescaled boundary-normal coordinate $\chi$ within the tomographic boundary layer according to Eq.~\eqref{eq:defchi} and express the distribution function (and each of the macroscopic variables) as a sum of the bulk solution and a tomographic boundary layer correction:
\begin{align}
    \Psi=\Psi_B + \Psi_T , \label{eq:hBT_sum}
\end{align}
where the latter is appreciable only near the device edges and $\Psi$ denotes any of the dependent variables. Asymptotic matching between the tomographic boundary layer and the bulk requires
\begin{align}
    \lim_{\chi\to\infty} \chi^a \Psi_T(\chi) = 0, \qquad a \in \mathbb{N}_0 ,
    \label{eq:matching}
\end{align}
for any non-negative integer $a$. Interestingly, the asymptotic expansion in powers of $\sqrt{k_e}$ employed here resembles that used in Ref.~\citep{liu20} to study slightly rarefied acoustic classical gas flows, where a ``viscous'' boundary layer emerges whose width scales with the square root of the Knudsen number.

The starting point of our analysis is the kinetic equation~\eqref{eq:BTE} for the electron distribution function. Substituting the decomposition in Eq.~\eqref{eq:hBT_sum} gives the governing equation for the bulk distribution function:
\begin{widetext}
\begin{align}
    v_i \frac{\partial h_B}{\partial x_i} +2 k_e E_i v_i-\frac{1}{r_c}v_i^\perp \frac{\partial h_B}{\partial v_i} = &
    - \frac{1}{k_e} \left(\left[h_B\right]_e-\delta \mu_B\right) 
    - \frac{1}{k_o} J_{\rm odd}\left([h_B]_o\right)
    - \frac{k_e}{G^2}\left(h_B-\delta \mu_B\right) , \label{eq:bulkBTE}
\end{align}
and its tomographic boundary layer correction:
\begin{align}
    v_it_it_j \frac{\partial h_T}{\partial x_j}+\frac{v_in_i}{\sqrt{k_ek_o}} \frac{\partial h_T}{\partial \chi} -\frac{1}{r_c}v_i^\perp \frac{\partial h_T}{\partial v_i} = &
    - \frac{1}{k_e} \left(\left[h_T\right]_e-\delta \mu_T\right) 
    - \frac{1}{k_o} J_{\rm odd}\left([h_T]_o\right) 
    - \frac{k_e}{G^2}\left(h_T-\delta \mu_T\right) . \label{eq:tomoBTE}
\end{align}
\end{widetext}
Note that, in the second equation, we have written the advective derivative in terms of the tangential coordinate and the rescaled boundary-normal coordinate~\eqref{eq:defchi}. The diffuse reflection condition at device edges becomes, for velocities satisfying \mbox{$v_in_i>0$},
\begin{align}
    \bigl[ h_B+h_T\bigr]_S = -\frac{1}{2}\int_{v_in_i<0} d\theta \, (v_in_i)\bigl[h_B+h_T\bigr]_S . \label{eq:diffuse}
\end{align}
The remainder of this section is structured as follows: The solution for the bulk distribution function in the limit~\mbox{$k_e\ll 1$} is given in Sec.~\ref{sec:bulk}. This analysis yields the bulk governing equations for the macroscopic variables in Eq.~\eqref{eq:bulkEqs} in Table~\ref{tbl:BCs_tomographicCorrections}. The solution of Eq.~\eqref{eq:tomoBTE} for the tomographic boundary layer correction in this limit is reported in Sec.~\ref{sec:tomographic}. This analysis gives the boundary conditions for the bulk governing equations [Eq.~\eqref{eq:bulkBCs} in Table~\ref{tbl:BCs_tomographicCorrections}] as well as corrections to the distribution function and the macroscopic variables within the tomographic boundary layer [Eq.~\eqref{eq:tomoCorrections} in Table~\ref{tbl:BCs_tomographicCorrections}]. The results of the calculations in this section are listed in Table~\ref{tbl:BCs_tomographicCorrections}, and their properties are discussed in the previous section.

\subsection{Bulk region}\label{sec:bulk}

The derivation of the bulk governing equations for the density and current~\eqref{eq:bulkEqs} as well as the bulk distribution function~\eqref{eq:hBulk} is comparatively straightforward and follows from a direct regular perturbative expansion in the square root of the even-mode Knudsen number $\sqrt{k_e}$. To this end, we expand the distribution function~$h_B$ in a power series in $\sqrt{k_e}$; cf.~Eq.~\eqref{eq:expansionhB}. Substituting this expansion into Eq.~\eqref{eq:bulkBTE} and collecting powers of~$\sqrt{k_e}$ gives
\begin{widetext}
\begin{subequations}
\begin{align}
	\bigl[h^{(0)}_B\bigr]_e = & \delta\mu^{(0)}_B , \label{eq:GE0_Lrc} \\[1ex]
	\bigl[h^{(1)}_B\bigr]_e = & \delta\mu^{(1)}_B , \\[1ex]
	\bigl[h^{(2)}_B\bigr]_e = & \delta\mu^{(2)}_B -v_i \frac{\partial h^{(0)}_B}{\partial x_i} +\frac{1}{r_c} v_k^\perp \frac{\partial h^{(0)}_B}{\partial v_k} - \frac{1}{k_o}J_\text{odd}(h_B^{(0)}) , \label{eq:GE1_Lrc}\\[1ex]
	\bigl[h^{(3)}_B\bigr]_e = & \delta\mu^{(3)}_B -v_i \frac{\partial h^{(1)}_B}{\partial x_i} +\frac{1}{r_c} v_k^\perp \frac{\partial h^{(1)}_B}{\partial v_k} - \frac{1}{k_o} J_\text{odd}(h_B^{(1)}), \\[1ex]
	\bigl[h^{(4)}_B\bigr]_e = & \delta\mu^{(4)}_B -2 E_i v_i - v_i \frac{\partial h^{(2)}_B}{\partial x_i} +\frac{1}{r_c} v_k^\perp \frac{\partial h^{(2)}_B}{\partial v_k} - \frac{1}{k_o}J_\text{odd}(h_B^{(2)}) - \frac{1}{G^2}\left(h^{(0)}_B - \delta\mu^{(0)}_B \right) , \label{eq:GE2_Lrc} 
    \end{align}
and, for higher orders \mbox{$n \geq 5$},
    \begin{align}
	\bigl[h^{(n)}_B\bigr]_e = & \delta\mu^{(n)}_B - v_i \frac{\partial h^{(n-2)}_B}{\partial x_i} +\frac{1}{r_c} v_k^\perp \frac{\partial h^{(n-2)}_B}{\partial v_k} - \frac{1}{k_o}J_\text{odd}(h_B^{(n-2)}) - \frac{1}{G^2}\left(h^{(n-4)}_B - \delta\mu^{(n-4)}_B\right) .
    \label{eq:GEn_Lrc}
\end{align}\label{eq:bulkGE_h}
\end{subequations}
\end{widetext}
These equations are solved sequentially and self-consistently by equating the odd and even components of the left- and right-hand sides of each equation in Eqs.~\eqref{eq:GE0_Lrc}--\eqref{eq:GEn_Lrc}. First, by definition, the density and current components of $h_B^{(n)}$ are set by $\delta \mu_B^{(n)}$ and $u_{B|i}^{(n)}$, respectively; cf.~Eqs.~\eqref{eq:momentsdensity} and~\eqref{eq:momentsvelocity}. Taking the first moment of the right-hand side of the bulk equations~\eqref{eq:bulkGE_h} at the $n$th order and equating it to $u_{B|i}^{(n)}$ for self-consistency then gives a constraint equation on \mbox{$\partial \delta \mu_B^{(n-2)}/\partial x_i$} that describes bulk momentum transport. These equations are exactly the Stokes-Ohm equations with higher-derivative corrections listed in Eq.~\eqref{eq:bulkEqs} in Table~\ref{tbl:BCs_tomographicCorrections}. The remaining odd-parity components on the right-hand side at the $n$th order determine the odd-parity part of $[h_B^{(n-2)}]_o$ beyond the current component \mbox{$2v_i u_{B|i}^{(n-2)}$}. These additional odd-parity corrections arise for $[h_B^{(2)}]_o$ and higher orders; they correspond to the terms in square brackets in Eq.~\eqref{eq:hBulk_n23} in Table~\ref{tbl:hSummary}. Next, in the same way, taking the zeroth moment of the right-hand sides gives a further constraint equation on the divergence of the velocity $u_{B|i}^{(n)}$, which at all orders is just the incompressibility condition listed in Eq.~\eqref{eq:bulkEqn_a} in Table~\ref{tbl:BCs_tomographicCorrections}. The remaining even-parity part then sets the even-parity component $[h_B^{(n)}]_e$ beyond the zeroth order. Again, such a correction arises at second order and higher, where it sets the third term in Eq.~\eqref{eq:hBulk_n23} in Table~\ref{tbl:hSummary}. This procedure must be performed to sixth order (\mbox{$n=6$}) to obtain the bulk governing equations listed in Table~\ref{tbl:BCs_tomographicCorrections} and the bulk distribution function in Table~\ref{tbl:hSummary}. 

The presence of the higher-moment correction to the distribution function in Eq.~\eqref{eq:hBulk_n23} immediately implies more complex slip boundary conditions on the velocity as well as the presence of a boundary layer: While the simple hydrodynamic deformations in Eq.~\eqref{eq:hBulk_n01} satisfy the diffuse boundary conditions at the device edges if no-slip boundary conditions (i.e., \mbox{$u_{B|i}^{(n)} t_i = 0$}) are applied, the same diffuse reflection condition cannot be satisfied by the bulk expansion at ${\it O}(k_e)$ and subsequent orders, due to the terms in the square bracket in Eq.~\eqref{eq:hBulk_n23}. These terms involve higher derivatives of the velocity, and setting them to zero in addition to a no-slip boundary condition would overdetermine the boundary conditions of the bulk equations. A tomographic boundary layer correction must be added to the bulk distribution function at~${\it O}(k_e)$ and subsequent orders to satisfy the diffuse reflection condition at these orders. The analysis of these corrections, which also yields the modified slip boundary conditions for the bulk equations, is described next.

\subsection{Tomographic boundary layer}\label{sec:tomographic}

The analysis of the tomographic boundary layer correction to the distribution function and the macroscopic variables is conducted by expanding each of the dependent variables in powers of $\sqrt{k_e}$:
\begin{align}
    \Psi_T = k_e \Psi_T^{(2)} + k_e^{3/2} \Psi_T^{(3)} + \cdots , \label{eq:keExpansionTomo}
\end{align}
where \mbox{$\Psi \in \{h, \delta \mu, u_i\}$}. Substituting Eq.~\eqref{eq:keExpansionTomo} into Eq.~\eqref{eq:tomoBTE} and collecting powers of $\sqrt{k_e}$ gives
\begin{widetext}
\begin{subequations}
\begin{align}
    \bigl[h_T^{(2)}\bigr]_e = &\delta \mu_T^{(2)} , \label{eq:tomoEqn2}\\[1ex]
	\bigl[h_T^{(3)}\bigr]_e = &\delta \mu_T^{(3)} - \frac{v_in_i}{\sqrt{k_o}} \frac{\partial h_T^{(2)}}{\partial \chi} , \label{eq:tomoEqn3}\\[1ex]
    \bigl[h_T^{(4)}\bigr]_e = &\delta \mu_T^{(4)} - \frac{v_in_i}{\sqrt{k_o}} \frac{\partial h_T^{(3)}}{\partial \chi} +\frac{1}{r_c} v_k^\perp 
    \frac{\partial h_T^{(2)}}{\partial v_k} 
    - v_it_i t_j\frac{\partial h_T^{(2)}}{\partial x_j} -\frac{1}{k_o}J_\text{odd}(h_T^{(2)}) , \label{eq:tomoEqn4}
\end{align}\label{eq:tomoGEn}
\end{subequations}
\end{widetext}
and so on. Next, the governing equations in Eq.~\eqref{eq:tomoGEn} are solved sequentially up to the same order %(i.e., ${\it O}(k_e)$)
as the bulk solution [i.e., ${\it O}(k_e)$].

First, Eq.~\eqref{eq:tomoEqn2} implies that $h_T^{(2)}$ has only an even-parity part in the zeroth moment, and any additional part that is not constrained by Eq.~\eqref{eq:tomoEqn2} must have odd parity.

Second, equating the odd-parity components of the left- and right-hand sides in Eq.~\eqref{eq:tomoEqn3} then gives
\begin{align}
    v_in_i \frac{\partial \delta \mu_T^{(2)}}{\partial \chi} = 0.
\end{align}
This, combined with the matching condition between the tomographic and bulk regions in Eq.~\eqref{eq:matching}, gives
\begin{align}
    \delta \mu_T^{(2)} = 0 , \label{eq:deltamuT2}
\end{align}
which is stated in Eq.~\eqref{eq:tomoLayerCorrection_mu2} in Table~\ref{tbl:BCs_tomographicCorrections}. There is, thus, no local deviation from the bulk electrochemical potential at any point in the boundary layer at this order. In turn, again from Eq.~\eqref{eq:tomoEqn2}, Eq.~\eqref{eq:deltamuT2} implies that the leading-order tomographic boundary layer correction to the distribution function is odd everywhere, i.e., 
\begin{align}
   [h_T^{(2)}]_e=0 . \label{eq:hT2isodd}
\end{align}
Thus, the general solution of Eq.~\eqref{eq:tomoEqn3} is
\begin{align}
    h_T^{(3)} = \delta \mu_T^{(3)}-\frac{v_in_i}{\sqrt{k_o}}\frac{\partial h_T^{(2)}}{\partial \chi} + A_T^{(3)} \label{eq:tomoSoln3}
\end{align}
with some odd function $A_T^{(3)}$ that will be constrained by a higher-order solution. Since by definition the zeroth moment of $h_T^{(3)}$ is equal to $\delta \mu_T^{(3)}$, the angle average of the second term in Eq.~\eqref{eq:tomoSoln3} must vanish. This gives
\begin{align}
    \frac{\partial (u_{T|i}^{(2)}n_i)}{\partial \chi} = 0,
\end{align}
and, thus, using the matching condition in Eq.~\eqref{eq:matching},
\begin{align}
    u_{T|i}^{(2)}n_i = 0 .
\end{align}
This equation is stated in Eq.~\eqref{eq:nopenetrationT} in Table~\ref{tbl:BCs_tomographicCorrections}. The overall no-penetration boundary condition then implies the separate no-penetration boundary condition on the bulk velocity profile $u_{B|i}^{(2)}$ stated in Eq.~\eqref{eq:nopenetrationbc} in Table~\ref{tbl:BCs_tomographicCorrections}.

Third, we proceed by analyzing Eq.~\eqref{eq:tomoEqn4}, which determines the odd component $[h_{T}^{(2)}]_o$. Substituting the form in Eq.~\eqref{eq:tomoSoln3} into Eq.~\eqref{eq:tomoEqn4} and equating the odd component of the left- and right-hand sides gives a governing equation for $h_T^{(2)}$, which is precisely the tomographic equation reported in Eq.~\eqref{eq:GEtomo2} in 
Table~\ref{tbl:hSummary}. For a practical solution of this equation, we expand the velocities at the boundary in an angular parameter $\theta$, measured from the tangential vector:
\begin{subequations}
\begin{align}
    v_i &= t_i \cos \theta + n_i \sin \theta ,\\[0.5ex]
    v_i^\perp &= t_i \sin \theta - n_i \cos \theta .
\end{align}
\end{subequations}
The tomographic equation~\eqref{eq:GEtomo2} in coordinate form then becomes
\begin{align}
    &\sin^2\theta \frac{\partial^2 h_T^{(2)}}{\partial \chi^2} -\frac{k_o}{r_c} \frac{\partial h_T^{(2)}}{\partial \theta}  \nonumber \\& = J_\text{odd}(h_T^{(2)}) + \sqrt{k_o} \sin\theta \frac{\partial \delta \mu_T^{(3)}}{\partial \chi} . \label{eq:GEtomo2theta}
\end{align}
The tomographic equation~\eqref{eq:GEtomo2theta} must be solved with the diffuse boundary condition~\eqref{eq:diffuse}, which links the tomographic layer and the bulk solutions at this order. Using the explicit form of the bulk solution $h_B^{(2)}$  [Eq.~\eqref{eq:hBulk_n23}], the requirement that the distribution is isotropic for \mbox{$n_i v_i \geq 0$} (corresponding to angles \mbox{$0\leq\theta<\pi$}) gives in coordinate form
\begin{widetext}
\begin{align}
    &\Bigl[ h_T^{(2)} \Bigr]_S = -2\cos\theta \, \Bigl[t_i u_{B|i}^{(2)} \Bigr]_S - \sin (2|\theta|) \Bigl[n_it_j S_{B|ij}^{(0)} \Bigr]_S
    + \kappa \frac{k_o}{1+(3k_o/r_c)^2} \left(\cos(3\theta)+\frac{3k_o}{r_c}\sin(3\theta)\right) \Bigl[n_it_j S_{B|ij}^{(0)} \Bigr]_S \nonumber \\[1ex]
    & 
    - \frac{k_o/2}{1+(3k_o/r_c)^2} \left(\cos(3\theta)+\frac{3k_o}{r_c}\sin(3\theta)\right) \biggl[t_in_jn_k \frac{\partial S_{B|ij}^{(0)}}{\partial x_k} \biggr]_S
    - \frac{3k_o/4}{1+(3k_o/r_c)^2} \left(\sin(3\theta)-\frac{3k_o}{r_c}\cos(3\theta)\right) \biggl[ n_in_jn_k \frac{\partial S_{B|ij}^{(0)}}{\partial x_k} \biggr]_S , \label{eq:hT2_eta0}
\end{align}
\end{widetext}
where we used that \mbox{$\delta \mu_T^{(2)} = 0$} [Eq.~\eqref{eq:deltamuT2}] as well as the no-penetration condition \mbox{$n_i u_{B|i}^{(2)} = 0$} and we made use of the antisymmetry of $h_T^{(2)}$ to extend the function to the full domain \mbox{$-\pi\leq \theta<\pi$}. Some derivative identities used to obtain the coordinate representation in Eq.~\eqref{eq:hT2_eta0} are listed in Appendix~\ref{app:derivatives}.

The solution of Eq.~\eqref{eq:GEtomo2theta} subject to the boundary condition in Eq.~\eqref{eq:hT2_eta0} gives, first, the velocity slip boundary conditions for the bulk equations, and, second, the tomographic boundary layer variables at order ${\it O}(k_e)$. The latter solution includes the solution for the boundary layer distribution function $h_T^{(2)}$, each of its moments, and the solution for the Hall field in the tomographic boundary layer at ${\it O}(k_e)$, i.e., \mbox{$\partial \delta \mu_T^{(3)}/\partial \chi$} [note that the derivative with respect to the rescaled boundary layer variable $\chi$ reduces the order by a power of $\sqrt{k_e}$]. Unlike the Knudsen layer analysis for the dual-relaxation-time collision operator~\cite{benshachar25a,benshachar25b}, where the slip conditions and kinetic corrections must be determined simultaneously, here we find that the boundary conditions for the bulk flow are independent of the tomographic boundary layer corrections. This is explained next.

To obtain the boundary conditions for the bulk flow, we project Eq.~\eqref{eq:GEtomo2theta} onto the tangential velocity component; i.e., we multiply  Eq.~\eqref{eq:GEtomo2theta} by $\cos\theta$ and integrate over \mbox{$-\pi\leq \theta \leq \pi$}, which gives
\begin{align}
    \frac{\partial^2}{\partial \chi^2}\int_{-\pi}^\pi d\theta \left(\cos\theta - \cos^3\theta\right)h_T^{(2)}  = 0  
    . \label{eq:obtainBCtomo}
\end{align}
Since all tomographic corrections must vanish for \mbox{$\chi\to \infty$} [cf.~Eq.~\eqref{eq:matching}], this implies
\begin{align}
    \int_{-\pi}^\pi d\theta \left(\cos\theta - \cos^3\theta\right)h_T^{(2)} = 0 \label{eq:obtainBCtomo2}
\end{align}
identically for all positions \mbox{$\chi\geq 0$} in the boundary layer. Substituting the boundary condition for $h_T^{(2)}$ [Eq.~\eqref{eq:hT2_eta0}] into Eq.~\eqref{eq:obtainBCtomo2} then gives the velocity slip boundary conditions at order ${\it O}(k_e)$, which are listed in Eq.~\eqref{eq:bulkBCslip} in Table~\ref{tbl:BCs_tomographicCorrections}. We note that the slip-length contribution [first term in Eq.~\eqref{eq:bulkBCslip}] arises from the $\sin(2|\theta|)$ term in Eq.~\eqref{eq:hT2_eta0}, while the explicit dependence on the odd-mode Knudsen number, magnetic field strength, boundary curvature, and derivatives of the stress [terms in round brackets in Eq.~\eqref{eq:bulkBCslip}] arises from the $\cos(3\theta)$ term in Eq.~\eqref{eq:hT2_eta0}. This procedure generates the slip boundary conditions in terms of exact constants. This is in contrast to the Knudsen layer analysis of hydrodynamic electron flow~\cite{benshachar25a,benshachar25b,raichev22}, where the slip boundary conditions must be written in terms of constants that are obtained through a numerical solution of integral equations.

Having obtained the slip boundary conditions for the bulk equations, we now turn to evaluating the tomographic boundary layer corrections for the distribution function, velocity profile, and Hall field at ${\it O}(k_e)$ (i.e., $h_T^{(2)}$, $u_{T|i}^{(2)}t_i$, and $\partial \delta \mu_T^{(3)}/\partial \chi$, respectively). First, $h_T^{(2)}$ solves the tomographic equation~\eqref{eq:GEtomo2theta} with boundary condition~\eqref{eq:hT2_eta0}. After substituting the slip boundary condition for $[t_i u_{B|i}^{(2)}]_S$ [Eq.~\eqref{eq:bulkBCslip}], Eq.~\eqref{eq:hT2_eta0} depends on the three linearly independent boundary expressions $[n_it_j S_{B|ij}^{(0)} ]_S$, $[t_in_jn_k\frac{\partial S_{B|ij}^{(0)}}{\partial x_k}]_S$, and $[n_in_jn_k\frac{\partial S_{B|ij}^{(0)}}{\partial x_k}]_S$, which suggests the form for the boundary layer correction $h_T^{(2)}$ stated in Eq.~\eqref{eq:hT_form} in Table~\ref{tbl:hSummary}. This form depends on three boundary-layer functions for the distribution function ($f_0$, $f_1$, and $f_2$), which are functions of the dimensionless boundary-normal coordinate~$\chi$ [Eq.~\eqref{eq:defchi}], the electron velocity~$v_i$, a dimensionless magnetic-field scaling parameter~$k_o/r_c$, and the relaxation rates of all odd modes $\lambda_5,\lambda_7,\ldots$. The numerical solution of the tomographic equation for these boundary-layer functions using a truncated moment expansion is discussed in Sec.~\ref{sec:truncatedMomentExp}. Likewise, $u_{T|i}^{(2)}t_i$ and \mbox{$\partial \delta \mu_T^{(3)}/\partial \chi$} are expressed in the same form in Eqs.~\eqref{eq:tomoLayerCorrection_u} and~\eqref{eq:tomoLayerCorrection_mu} in Table~\ref{tbl:BCs_tomographicCorrections}, with corresponding tomographic boundary-layer functions $\mathcal{Y}_0,\mathcal{Y}_1$, and $\mathcal{Y}_2$ for the velocity profile and $\mathcal{T}_0,\mathcal{T}_1$, and $\mathcal{T}_2$ for the Hall field. These are obtained from the solution of the tomographic boundary layer deformation by 
\begin{subequations}
\begin{align}
    \mathcal{Y}_i &= \int \frac{d\theta}{2\pi} \cos \theta \, f_i , \label{eq:definition_Y}\\
    \mathcal{T}_i &= - \frac{1}{2} \frac{\partial^2}{\partial \chi^2} \int \frac{d\theta}{2\pi} \sin 3 \theta \, f_i ,\label{eq:definition_T}
\end{align}
\end{subequations}
where the second identity follows from a projection of the tomographic equation~\eqref{eq:GEtomo2theta} onto the normal velocity component. Results for these tomographic boundary-layer functions are reported in Figs.~\ref{fig:tomoFun} and~\ref{fig:tomoFun2} for the relaxation rates given by Eq.~\eqref{eq:lambda_m_model} with \mbox{$p=0$} and in Appendix~\ref{app:p4_tomoFun} for~\mbox{$p=4$}.

\subsection{Numerical solution of the tomographic boundary-layer functions}\label{sec:truncatedMomentExp}

We now present an efficient numerical solution of Eq.~\eqref{eq:GEtomo2} using a truncated moment expansion. Substituting Eq.~\eqref{eq:hT_form}---which expresses the tomographic boundary deformation $h_T^{(2)}$ in terms of three scalar boundary-layer functions $f_0, f_1$, and $f_2$---into the tomographic equation~\eqref{eq:GEtomo2}, we obtain the governing equations for each $f_i$, $\mathcal{Y}_i$, and $\mathcal{T}_i$ [the latter are defined in Eqs.~\eqref{eq:definition_Y} and~\eqref{eq:definition_T}]:
\begin{align}
    &\sin^2\theta \frac{\partial^2 f_i}{\partial \chi^2} \nonumber -\frac{k_o}{r_c} \frac{\partial f_i}{\partial \theta} \\
    &\quad = J_\text{odd}(f_i) + \sin\theta \biggl(\mathcal{T}_i + \frac{2k_o}{r_c}\mathcal{Y}_i\biggr) , \label{eq:fi}
\end{align}
with \mbox{$i\in\{0,1,2\}$}. The boundary condition for $h_T^{(2)}$ in Eq.~\eqref{eq:hT2_eta0} gives the following condition for each $f_i$ at \mbox{$\chi=0$}:
\begin{subequations}
\begin{align}
    \bigl[ f_0 \bigr]_S = & \frac{64}{15\pi}\cos\theta - \sin(2|\theta|), \\[1ex]
    \bigl[ f_1 \bigr]_S = & \frac{\cos\theta+\cos(3\theta)+3k_o/r_c\sin(3\theta)}{1+(3k_o/r_c)^2}, \\[1ex]
    \bigl[ f_2 \bigr]_S = & \frac{3}{4}\frac{3k_o/r_c[\cos\theta+\cos(3\theta)]-\sin(3\theta)}{1+(3k_o/r_c)^2} .
\end{align}\label{eq:fi_BCs}%
\end{subequations}
We solve these equations using a truncated moment expansion: Each of the distribution functions is expanded in the truncated series,
\begin{align}
	f_i(\chi, \theta) = \sum_{m \, {\rm odd}}^N \bigl[a_m(\chi) \cos(m\theta) + b_m(\chi) \sin(m\theta)\bigr] ,
    \label{eq:tomoLayer_momentExp}
\end{align}
for each $i\in\{0,1,2\}$, where $a_m$ and $b_m$ are functions of $\chi$ and $k_o/r_c$, and $N$ is an integer that sets the cutoff of the expansion. Only odd $m\geq 1$ contribute to this expansion since the tomographic boundary-layer functions all have odd parity [cf.~Eq.~\eqref{eq:hT2isodd}]. Note that the no-penetration boundary condition implies \mbox{$b_1=0$} and Eq.~\eqref{eq:obtainBCtomo2} implies \mbox{$a_1=a_3$}. Substituting Eq.~\eqref{eq:tomoLayer_momentExp} into Eq.~\eqref{eq:fi} and then projecting Eq.~\eqref{eq:fi} onto the subspace spanned by the basis functions of Eq.~\eqref{eq:tomoLayer_momentExp} gives a system of coupled ordinary differential equations for $a_m$ and $b_m$:
\begin{align}
	\begin{pmatrix}
		A_N	&	0 \\[1ex]
		0	&	B_N
	\end{pmatrix}
	\begin{pmatrix}
		\bf{a} \\[1ex]	\bf{b}
	\end{pmatrix}''
	=\begin{pmatrix}
		D_N					&	\frac{k_o}{r_c} C_N \\[1ex]
		- \frac{k_o}{r_c} C_N		&	D_N
	\end{pmatrix}
	\begin{pmatrix}
		\bf{a} \\[1ex]	\bf{b}
	\end{pmatrix}. \label{eq:anbn_eqn}
\end{align}
Here, $\mathbf{a}=(a_3,\ldots, a_{N})^T,\mathbf{b}=(b_3,\ldots, b_{N})^T$, $D_N=\text{diag}\{1,\lambda_5,  \lambda_7, \ldots\}$, the prime symbol is used to denote differentiation with respect to $\chi$, $C_N$ is a diagonal matrix with entries $[C_N]_{kk'}=(2k+1) \delta_{kk'}$, and we define
\begin{subequations}
\begin{align}
	A_N = & 
	\begin{pmatrix}
		\frac{1}{4} 		&	-\frac{1}{4}		&	0	&	0		&	\cdots	 &	0 \\[1ex]
		-\frac{1}{4}		&	\frac{1}{2}		&	-\frac{1}{4}		&	0	&	\cdots	&	0 \\[1ex]
		0		&	-\frac{1}{4}		&	\frac{1}{2}	&	-\frac{1}{4}		&	\cdots	&	0 \\[1ex]
		\vdots	&			&		&	\ddots&		&		\vdots \\[1ex]
		0		&	0		&	0	&		-\frac{1}{4}	&	\frac{1}{2}	&	-\frac{1}{4} \\[1ex]
		0		&	0		&	0	&		0	&	-\frac{1}{4}	&	\frac{1}{2}
	\end{pmatrix} , \\[2ex]
	B_N = & 
	\begin{pmatrix}
		\frac{1}{2} 		&	-\frac{1}{4}		&	0	&	0	&		\cdots	 &	0 \\[1ex]
		-\frac{1}{4}		&	\frac{1}{2}		&	-\frac{1}{4}	&		0	&	\cdots	&	0 \\[1ex]
		0		&	-\frac{1}{4}		&	\frac{1}{2}	&	-\frac{1}{4}		&	\cdots	&	0 \\[1ex]
		\vdots	&			&		&	\ddots&		&	\vdots \\[1ex]
		0		&	0		&	0	&		-\frac{1}{4}	&	\frac{1}{2}	&	-\frac{1}{4} \\[1ex]
		0		&	0		&	0	&	0		&	-\frac{1}{4}	&	\frac{1}{2}
	\end{pmatrix} .
\end{align}
\end{subequations}
Equation~\eqref{eq:anbn_eqn} is then solved numerically for each fixed value of $k_o/r_c$ by expanding in eigenvectors 
\begin{align}
	\begin{pmatrix}
		\bf{a} \\[1ex]	\bf{b}
	\end{pmatrix}
	= \sum_j c_j {\bf v}_j \exp \Bigl(-\sqrt{\lambda_j} \, \chi \Bigr), \label{eq:expansion}
\end{align}
where ${\bf v}_j$ and $\lambda_j$ are the eigenvectors and eigenvalues, respectively, of the matrix
\begin{align}
	M_N =   \begin{pmatrix}
		A_N	&	0 \\[1ex]
		0	&	B_N
	\end{pmatrix}^{-1} \begin{pmatrix}
		D_N					&	(k_o/r_c) C_N \\[1ex]
		-(k_o/r_c) C_N		&	D_N
	\end{pmatrix} . \label{eq:MN}
\end{align}
The coefficients $c_j$ in Eq.~\eqref{eq:expansion} are then found by projecting the boundary conditions in Eq.~\eqref{eq:fi_BCs} onto the basis functions of Eq.~\eqref{eq:tomoLayer_momentExp}, which gives boundary conditions for $\bf{a}$ and $\bf{b}$ at \mbox{$\chi=0$}. They are, for $f_0$,
\begin{subequations}
\begin{align}
    [a_m]_S &= 
    \dfrac{8}{\pi(m+2)(m-2)} , 
    \\[1.5ex]
    [b_m]_S &=0 ;
\end{align}
\end{subequations}
for $f_1$,
\begin{subequations}
\begin{align}
    [a_m]_S &=  \begin{cases}
    \dfrac{1}{1+(3k_o/r_c)^2} & m=3 , \\[1ex]
    0 & {\rm otherwise} ,
    \end{cases} \\[1.5ex]
    [b_m]_S &= \begin{cases}
    \dfrac{3 k_o/r_c}{1+(3k_o/r_c)^2} & m=3 , \\[1ex]
    0 & {\rm otherwise} ;
    \end{cases} \label{eq:bcf1bn}
\end{align}
\end{subequations}
and, for $f_2$,
\begin{subequations}
\begin{align}
    [a_m]_S &= \begin{cases}
    \dfrac{9}{4} \dfrac{k_o/r_c}{1+(3k_o/r_c)^2} & m=3 , \\[1.5ex]
    0 & {\rm otherwise} ,
    \end{cases}  \\[1.5ex]
    [b_m]_S &= \begin{cases}
    - \dfrac{3}{4} \dfrac{1}{1+(3k_o/r_c)^2} & m=3 , \\[1ex]
    0 & {\rm otherwise} .
    \end{cases} \label{eq:bcf2bn}
\end{align}
\end{subequations}
The solutions for ${\cal Y}_i$ and ${\cal T}_i$ are then obtained from
\begin{subequations}
\begin{align}
    {\cal Y}_i &= \frac{a_1}{2} = \frac{a_3}{2} , \\
    {\cal T}_i &= - \frac{1}{4} \frac{d^2 b_3}{d \chi^2} . \label{eq:TiDefnApp}
\end{align}
\end{subequations}
These expressions follow by substituting Eq.~\eqref{eq:tomoLayer_momentExp} into Eqs.~\eqref{eq:definition_Y} and~\eqref{eq:definition_T}. The value of $N$ is increased until convergence is obtained. A final value of~\mbox{$N=100$} was used in generating the plots of the tomographic boundary-layer functions in Fig.~\ref{fig:tomoFun}.

\section{An example: Channel flow} \label{sec:channel_flow}

As an application of the framework derived above, we now present the asymptotic solution for flow in a channel. We choose the width of the channel as the macroscopic length scale $L$, so that in terms of the rescaled coordinate the diffuse edges lie at \mbox{$x=\pm 1/2$}. The electric field is applied in the negative $y$ direction with \mbox{$E_y=-1$} (such that electron flow is in the positive $y$ direction), and the perpendicular magnetic field is oriented along the $z$ direction. For this geometry, the velocity profile $u_i$ is oriented along the $y$ direction, and the Hall field $-\partial \delta \mu/\partial x_i$ is oriented along the~$x$ direction. Both quantities depend only on the transverse coordinate~$x$, which implies that the tomographic Stokes-Ohm equations~\eqref{eq:bulkEqs} for the velocity profile and the Hall field decouple. The asymptotic solution for the flow velocity and Hall field up to ${\it O}(k_e)$ is presented next. This solution reveals that slip and tomographic-layer corrections are required to obtain an accurate description of these profiles, which we verify by comparing with the direct numerical solution of the kinetic equation~\eqref{eq:BTE}.

\subsection{Velocity profile}

To obtain the asymptotic solution for electron flow in a channel, we proceed as follows: First, the incompressibility condition~\eqref{eq:bulkEqn_a} ensures that the flow velocity is directed along the channel only (i.e., \mbox{$u_{B|x}^{(n)}=0$} for all $n$). The equation for the leading-order bulk velocity $u_{B|y}^{(0)}$ follows from Eq.~\eqref{eq:bulkEqn_c}, which takes the Stokes-Ohm form
\begin{align}
    \frac{1}{4} \frac{\partial^2 u_{B|y}^{(0)}}{\partial x^2} - \frac{u^{(0)}_{B|y}}{G^2} &= -1 \label{eq:LOchannel}
\end{align}
and which must be solved with no-slip boundary conditions~\eqref{eq:bulkBCns} at the channel edges, \mbox{$u_{B|y}^{(0)}(x=\pm1/2) = 0$}. This gives the hydrodynamic Poiseuille velocity profile~\citep{alekseev16,matthaiakakis2020,holder19} 
\begin{align}
    u_{B|y}^{(0)}(x) &= G^2 \biggl(1 - \frac{\cosh(2x/G)}{\cosh(1/G)}\biggr) . \label{eq:uy0}
\end{align}
Likewise, Eq.~\eqref{eq:bulkEqn_d} for the next-to-leading-order correction $u_{B|y}^{(1)}$ takes the same form as Eq.~\eqref{eq:LOchannel} but without the electric-field source term. The solution with the no-slip boundary condition~\eqref{eq:bulkBCns}, therefore, vanishes identically:
\begin{align}
    u_{B|y}^{(1)} &= 0 . \label{eq:uy1}
\end{align}
At second order ${\it O}(k_e)$, Eq.~\eqref{eq:bulkEqn_n2} for $u_{B|y}^{(2)}$ becomes
\begin{align}
\frac{1}{4} \frac{\partial^2 u_{B|y}^{(2)}}{\partial x^2} - \frac{u_{B|y}^{(2)}}{G^2} &= - \frac{1}{16} \frac{k_o}{1+(3k_o/r_c)^2} \frac{\partial^4 u_{B|y}^{(0)}}{\partial x^4} , \label{eq:uByeq}
\end{align}
where the leading-order solution~\eqref{eq:uy0} acts as a source term. The leading-order profile~\eqref{eq:uy0} also determines the modified slip boundary condition~\eqref{eq:bulkBCslip} for Eq.~\eqref{eq:uByeq}, which reads
\begin{align}
    \Bigl[u_{B|y}^{(2)} \Bigr]_S =   
         \frac{64}{15\pi} G \tanh(1/G)
         + \frac{k_o}{1+(3k_o/r_c)^2} , \label{eq:slipbcchannel}
\end{align}
where we used that \mbox{$\kappa = 0$} for a straight channel as well as
\begin{subequations}
\begin{align}
\Bigl[ n_i t_j S_{B|ij}^{(0)} \Bigr]_S &= \pm 2 G \tanh(1/G) , \label{eq:bcchannel1} \\
\biggl[ t_i n_j n_k \frac{\partial S_{B|ij}^{(0)}}{\partial x_k} \biggr]_S &= \mp 4, \label{eq:bcchannel2} \\
\biggl[ n_i n_j n_k \frac{\partial S_{B|ij}^{(0)}}{\partial x_k} \biggr]_S &= 0 \label{eq:bcchannel3} ,
\end{align}
\end{subequations}
where the upper sign holds at the left and the lower sign at the right channel boundary (recall that, in our convention, the tangential vector forms a right-handed screw with the normal vector, the latter of which points into the flow domain). The solution of Eq.~\eqref{eq:uByeq} subject to the boundary conditions in Eq.~\eqref{eq:slipbcchannel} is
\begin{widetext}
\begin{align}
    u_{B|y}^{(2)}(x) &=\biggl(\frac{64}{15\pi}G \tanh(1/G) + \frac{k_o}{1+(3k_o/r_c)^2} \biggr) \frac{\cosh(2x/G)}{\cosh(1/G)} \nonumber \\[1ex]
    &\qquad + \frac{k_o}{1+(3k_o/r_c)^2} \frac{1}{2G\cosh(1/G)} \Bigl(2x \sinh(2x/G)-\tanh(1/G)\cosh(2x/G)\Bigr) . 
    \label{eq:uy2}
\end{align}
The terms in the first line are due to the slip boundary condition~\eqref{eq:slipbcchannel}, and the terms in the second line, which vanish at the channel boundaries, are the bulk rarefaction (i.e., finite-wavelength) correction. Finally, at this order, there is also a tomographic boundary layer correction to the velocity profile, which follows from Eq.~\eqref{eq:tomoLayerCorrection_u} [using again Eqs.~\eqref{eq:bcchannel1}--\eqref{eq:bcchannel3}]: 
\begin{align}
    u_{T|y}^{(2)} &= - 2G \tanh(1/G) \left[\mathcal{Y}_0\left(\frac{\tfrac{1}{2}+x}{\sqrt{k_ek_o}};\frac{k_o}{r_c}\right)+\mathcal{Y}_0\left(\frac{\tfrac{1}{2}-x}{\sqrt{k_ek_o}};\frac{k_o}{r_c}\right)\right] -2 k_o \left[\mathcal{Y}_1\left(\frac{\tfrac{1}{2}+x}{\sqrt{k_ek_o}};\frac{k_o}{r_c}\right)+\mathcal{Y}_1\left(\frac{\tfrac{1}{2}-x}{\sqrt{k_ek_o}};\frac{k_o}{r_c}\right)\right] . \label{eq:uyT2}
\end{align}
\end{widetext}
Equations~\eqref{eq:uy0},~\eqref{eq:uy1},~\eqref{eq:uy2}, and~\eqref{eq:uyT2} dictate the asymptotic solution of the velocity profile up to ${\it O}(k_e)$, with the full solution reading
\begin{align}
    u_y = & u_{B|y}^{(0)} + k_e \left(u_{B|y}^{(2)} + u_{T|y}^{(2)}\right)+ {\it O}(k_e^{3/2}) . \label{eq:ux_channel} 
\end{align}

\begin{figure*}
    \centering
    \includegraphics{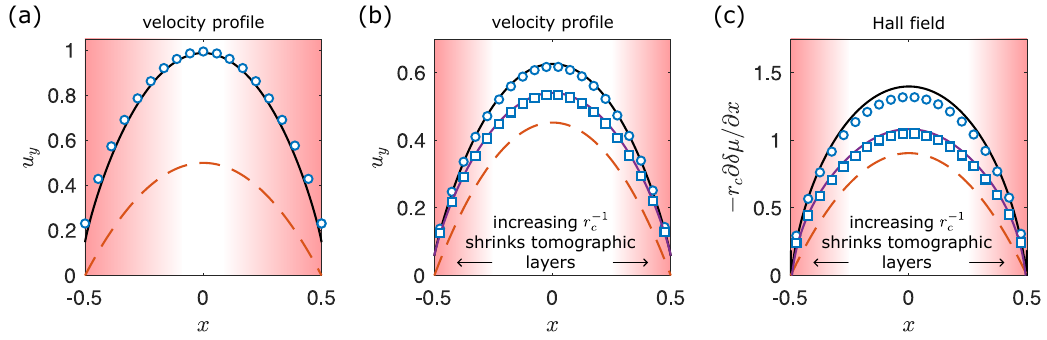}
    \caption{Velocity ($u_y$) and rescaled Hall field (\mbox{$-r_c\partial \delta \mu/\partial x$}) profiles in a channel for odd-mode relaxation rates~\eqref{eq:lambda_m_model} with $p=4$. Direct numerical solutions (blue open circles and squares) are compared with the asymptotic solution (solid black and purple lines) and the hydrodynamic solution (dashed red line). Parameter values are (a) \mbox{$k_e=0.2$}, \mbox{$k_o=2$}, \mbox{$G\to\infty$}, and $r_c^{-1}=0$; (b), (c) \mbox{$k_e=0.075$}, \mbox{$k_o=2$}, \mbox{$G=2$}, and \mbox{$r_c^{-1}=0.025$} (open circles, black solid line) and \mbox{$r_c^{-1}=1$} (open squares, purple solid line). The red shaded region shows the approximate extent of the tomographic boundary layer with arrows indicating the shrinking of this layer with increased magnetic field strength.}
    \label{fig:channelFlow}
\end{figure*}

Before analyzing the asymptotic expression~\eqref{eq:ux_channel} in detail, we compare 
this prediction with direct numerical solutions of the linearized Boltzmann equation~\eqref{eq:BTE} in a channel (see Fig.~\ref{fig:channelFlow}). In particular, we consider solutions for the (dimensionless) odd-mode relaxation rates given by Eq.~\eqref{eq:lambda_m_model} with~\mbox{$p=4$}. Details of the numerical solution are provided in Appendix~\ref{app:numSol}. We find excellent agreement between the asymptotic and numerical solutions for the velocity profile, even for a non-negligible even-mode Knudsen number \mbox{$k_e=0.2$}. Indeed, the velocity profile is predicted to within 10\% over the entire spatial domain for all parameter values shown, and similar agreement is found for other parameter values. This is to be compared with the hydrodynamic no-slip solution, for which the velocity profile is given by Eq.~\eqref{eq:uy0} (red dashed line in Fig.~\ref{fig:channelFlow}). The hydrodynamic prediction exhibits an error of approximately $100\%$ throughout the spatial domain. A systematic study of the error between the asymptotic predictions and the numerical solution as a function of~$k_e$ (see Appendix~\ref{app:ke_convergence}) shows a small error that scales as a fractional power~\mbox{$\sim k_e^{3/2}$} for small~$k_e$, as indeed expected from the truncation of the expansion in Eq.~\eqref{eq:keExpansion}. The calculation also shows that the channel conductance is predicted accurately up to relatively large Knudsen numbers; for example, for \mbox{$k_e<0.4$} the channel conductance is predicted to within \mbox{$12\%$} for representative parameter values and for \mbox{$k_e<0.2$} to within \mbox{$3\%$}. This shows that the asymptotic solution provides accurate predictions at small but finite even-mode Knudsen numbers~$k_e$. 

%++++++++++++++++++++++++++++++++++++++++
%++++++++++++++++++++++++++++++++++++++++

We now discuss key features of the asymptotic solution for the velocity profile in more detail. At leading order in $k_e$, the velocity profile in Eq.~\eqref{eq:ux_channel} is given by the widely reported hydrodynamic $\cosh$ profile with no-slip boundary conditions [see Eq.~\eqref{eq:uy0}]. At first order in $k_e$, however, the asymptotic solution strongly deviates from previous hydrodynamic solutions. In particular, at this order Eq.~\eqref{eq:ux_channel} exhibits (i) a significant velocity slip condition [the first line in Eq.~\eqref{eq:uy2}], which acts to increase the velocity profile; (ii) tomographic bulk corrections [the second line in Eq.~\eqref{eq:uy2}], which act to decrease the velocity profile; and (iii) tomographic boundary layer corrections [the $u_{T|y}^{(2)}$ term in Eq.~\eqref{eq:ux_channel}], which act to decrease the velocity profile. The tomographic boundary layer extends significantly into the flow domain (see the red shaded region in Fig.~\ref{fig:channelFlow}) relative to the extent of the Knudsen layer (i.e., the analogous nonhydrodynamic boundary layer in the absence of the tomographic effect~\citep{benshachar25a,benshachar25b}) and is, thus, essential for obtaining the correct velocity profile throughout the spatial domain. In the absence of an external magnetic field, the tomographic boundary layer extends over the scaled distance \mbox{$\sim \sqrt{k_ek_o}$}; see Fig.~\ref{fig:channelFlow}(a). However, the width of the tomographic boundary layer is dramatically reduced in the presence of an external magnetic field; see Fig.~\ref{fig:channelFlow}(b). In particular, when the cyclotron radius is smaller than the odd-mode mean free path (\mbox{$R_c<\ell_o$}), the scaled tomographic boundary layer width scales instead as \mbox{$\sim \sqrt{k_er_c}$}. Moreover, the magnitudes of the velocity slip in the bulk solution and the bulk finite-wavelength correction decrease significantly when an external magnetic field is applied. In particular, both decrease in magnitude as a Lorentzian with argument \mbox{$3\ell_o/R_c$}. Together, these phenomena induce a negative magnetoconductance, which can be seen in Fig.~\ref{fig:channelFlow}(b). The strong deviation between the asymptotic theory and the hydrodynamic result highlights that the distinct shape of the true velocity profile requires (i) higher-derivative corrections to the Stokes-Ohm equation, (ii) a nonzero and significant velocity slip condition, and (iii) tomographic-layer corrections to the velocity profile, all of which are rigorously captured by the asymptotic solution in Eq.~\eqref{eq:ux_channel}. 

\subsection{Hall field}\label{eq:hallfieldchannel}

At leading order, the Hall field in the bulk is obtained by solving Eq.~\eqref{eq:bulkEqn_b}:
\begin{align}
- \frac{1}{2} \frac{\partial \delta \mu_B^{(0)}}{\partial x} - \frac{1}{r_c} u^{(0)}_{B|y} &= 0 ,
\end{align}
which gives
\begin{align}
    - \frac{\partial \delta \mu_B^{(0)}}{\partial x} &= \frac{2}{r_c} u_{B|y}^{(0)}
    \label{eq:mu0}
\end{align}
with the leading-order bulk velocity profile reported in Eq.~\eqref{eq:uy0}. With the same constraint equation, Eq.~\eqref{eq:uy1} implies
\begin{align}
    - \frac{\partial \delta \mu_B^{(1)}}{\partial x} &= 0 .
\end{align}
Likewise, Eq.~\eqref{eq:bulkEqn_c} implies
\begin{align}
    - \frac{\partial \delta \mu_B^{(2)}}{\partial x} &= \frac{2}{r_c} u_{B|y}^{(2)} 
    \label{eq:mu2}
\end{align}
with the higher-order velocity profile~\eqref{eq:uy2} that involves finite-wavelength corrections and slip boundary conditions. On top of this, the bulk Hall field acquires corrections in the tomographic boundary layer at ${\it O}(k_e)$, which follow from the derivative of the third-order electrochemical potential \mbox{$\delta \mu_T^{(3)}$} with respect to the rescaled boundary layer coordinate [Eq.~\eqref{eq:tomoLayerCorrection_mu}]:
\begin{align}
    &-\sqrt{k_e}\frac{\partial \delta \mu_T^{(3)}}{\partial x} = \frac{2}{r_c}u_{T|y}^{(2)}\nonumber \\&\hspace{5em}-2 \left[\mathcal{T}_1\left(\frac{\tfrac{1}{2}+x}{\sqrt{k_ek_o}};\frac{k_o}{r_c}\right)+\mathcal{T}_1\left(\frac{\tfrac{1}{2}-x}{\sqrt{k_ek_o}};\frac{k_o}{r_c}\right)\right] \nonumber \\[1ex]
    &- \frac{2G}{k_o} \tanh(\tfrac{1}{G}) \left[\mathcal{T}_0\left(\frac{\tfrac{1}{2}+x}{\sqrt{k_ek_o}};\frac{k_o}{r_c}\right)+\mathcal{T}_0\left(\frac{\tfrac{1}{2}-x}{\sqrt{k_ek_o}};\frac{k_o}{r_c}\right)\right] 
    . \label{eq:dmudxT}
\end{align}
The factor $\sqrt{k_e}$ on the left-hand side in Eq.~\eqref{eq:dmudxT} arises from the ${\it O}(\sqrt{k_e})$ width of the tomographic boundary layer. The full asymptotic solution for the Hall field up to ${\it O}(k_e)$, thus, reads
\begin{align}
    -\frac{\partial \delta \mu}{\partial x} = & \frac{2}{r_c} \bigl(u_{B|y}^{(0)} + k_e u_{B|y}^{(2)}\bigr) + k_e \Big(- \sqrt{k_e}\frac{\partial \delta \mu_T^{(3)}}{\partial x}\Big) +  {\it o}(k_e),
    \label{eq:Ehall_channel}
\end{align}
where the second term is given by Eq.~\eqref{eq:dmudxT}. We note that, while the Hall field is well defined, integrating Eq.~\eqref{eq:Ehall_channel} defines the electrochemical potential in the channel up to an arbitrary integration constant. This is consistent with solutions to the kinetic equation~\eqref{eq:BTE} subject to diffuse reflection~\eqref{eq:diffuse} at sample boundaries being defined up to an arbitrary constant shift in $h$.

%++++++++++++++++++++++++++++++++++++++++
\begin{figure*}
    \centering
    \includegraphics{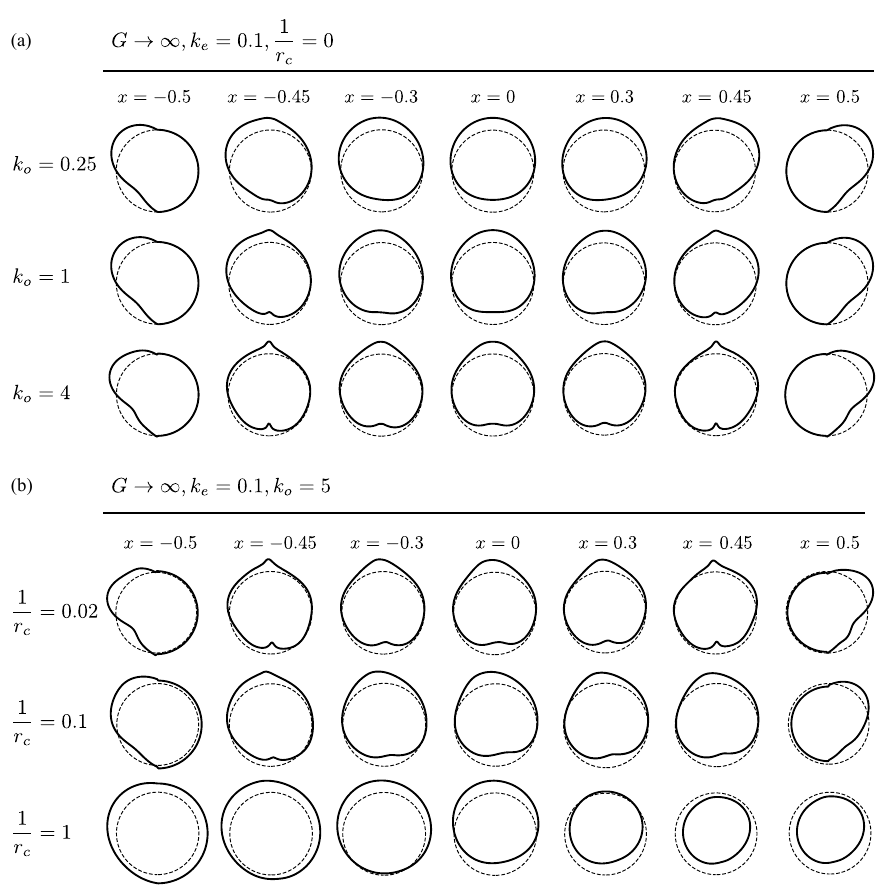}
    \caption{
    Microscopic Fermi surface deformation (solid black lines) for tomographic electron flow in a channel with $\lambda_m$ given by Eq.~\eqref{eq:lambda_m_model} with $p=0$ (a) for three different odd-parity mean free paths (top to bottom rows) without a magnetic field; and (b) for three different magnetic fields with a fixed odd-parity mean free path. Shown are the deformations at different positions in the channel at \mbox{$x=-0.5,-0.45,-0.3,0,0.3,0.45$}, and $0.5$ (left to right columns), where the channel boundaries are at \mbox{$x=\pm0.5$}. The dashed circles indicate the equilibrium Fermi surface. The diffuse boundary condition~\eqref{eq:diffuseintro}, where electrons impinging on the boundary are isotropically reflected, is directly apparent in the microscopic distribution at the boundary. In addition, all distributions have a center-of-mass displacement in the vertical direction, indicating electron flow in the positive $y$ direction. Without a magnetic field, the electron density is fixed across the channel, corresponding to an unchanged area of the distribution. The area changes in the presence of a magnetic field, corresponding to the accumulation of charge at channel boundaries and the buildup of a Hall field. Parameters in (a) are \mbox{$G\to \infty$} (no disorder), \mbox{$r_c^{-1}=0$} (no magnetic field), \mbox{$k_e=0.1$}, and \mbox{$k_o=0.25,1$}, and $4$ (top to bottom rows). The distribution at the boundary is strongly nonuniform and relaxes by collisions to a hydrodynamiclike form (i.e., a deformation that predominantly involves a center-of-mass shift of the equilibrium distribution) in the channel center. The range over which this happens defines the tomographic boundary layer. The extent of this layer is \mbox{${\it O}(\sqrt{k_ek_o})$}, and indeed it is apparent from (a) that it increases from the top row to the bottom row. Likewise, (b) shows the microscopic deformation with \mbox{$G\to \infty$}, \mbox{$k_e=0.1$}, and \mbox{$k_o=5$} for three different magnetic fields, \mbox{$r_c^{-1} = 0.02, 0.1$}, and $1$. While the tomographic layer extends to the channel center for weak fields, it is quickly suppressed at stronger fields, where it extends only over a range \mbox{${\it O}(\sqrt{k_er_c})$}.}
    \label{fig:channel_FermiSurface}
\end{figure*}
%++++++++++++++++++++++++++++++++++++++++

In the bulk of the channel, the Hall field~\eqref{eq:Ehall_channel} is given by the Lorentz force only up to ${\it O}(k_e)$. No Hall-viscosity contributions arise, since these are of ${\it O}(k_e^2)$. The leading-order deviation of the Hall field from this classical description is, thus, given by the tomographic boundary-layer functions $\mathcal{T}_0$ and $\mathcal{T}_1$. This gives rise to more rounded Hall-field profiles in Fig.~\ref{fig:channelFlow}(c) relative to the respective velocity profiles in Fig.~\ref{fig:channelFlow}(b).

At ${\it O}(k_e^2)$, the Hall field deviates from the Lorentz force prediction. In particular, Eq.~\eqref{eq:bulkEqs} dictates that the bulk Hall field ($-\partial \delta \mu_B/\partial x$) is related to the velocity profile via %\johannes{Should the first term be the bulk velocity, $u_y\to u_{B|y}$?}
\begin{align}
    -\frac{\partial \delta \mu_B}{\partial x}= & \frac{2}{r_c}\left[u_{B|y} + \frac{k_e^2}{2}\left(1+\frac{3k_o^2}{2G^2}\frac{1}{1+(3k_o/r_c)^2}\right)\frac{\partial^2 u_{B|y}}{\partial x^2}\right]\nonumber \\& + {\it o}(k_e^2) .\label{eq:channl_EHall_TomoEqn}
\end{align}
The first term in the square brackets is the (continuum) Lorentz force contribution, while the second term (with prefactor $k_e^2$) gives the leading-order rarefaction (i.e., finite-wavelength) correction to the bulk Hall field. This correction depends strongly on the magnetic field, where the first term in the round brackets is the usual Hall viscosity contribution~\citep{benshachar25a,benshachar25b,matthaiakakis2020}, while the second term---which is of the same order---is a new contribution associated with the tomographic effect. Thus, in this flow regime, the deviation of the Hall field from Lorentz-force predictions is not given by the Hall viscosity alone. Moreover, the tomographic contribution to Eq.~\eqref{eq:channl_EHall_TomoEqn} depends on the magnetic field strength, following a Lorentzian with argument~\mbox{$3 k_o/r_c = 3\ell_o/R_c$}, and is suppressed at moderate magnetic fields for which \mbox{$r_c \sim 3k_o$}, i.e., when the cyclotron radius is comparable to the large odd-mode mean free path. The suppression of the tomographic contribution to Eq.~\eqref{eq:channl_EHall_TomoEqn} gives a simple method for determining the odd-mode mean free path: For sufficiently large magnetic field strengths (i.e., for which \mbox{$R_c<\ell_o$}), the tomographic correction is suppressed and Eq.~\eqref{eq:channl_EHall_TomoEqn} recovers the hydrodynamic prediction, for which the contribution from the tomographic effect vanishes.

The curvature of the Hall field at the center of the channel is primarily dictated by the bulk solution. We, therefore, observe from Eq.~\eqref{eq:channl_EHall_TomoEqn} that the tomographic effect decreases the curvature of the Hall field at the center of the channel. This suggests that the tomographic effect is not responsible for the enhanced Hall-field curvature measured in Ref.~\citep{sulpizio19}. Moreover, these measurements were conducted at a significant magnetic field strength of \mbox{$1/r_c\approx 1.3$}, for which we would expect tomographic phenomena to be suppressed. We note that the analysis here assumes a smooth boundary geometry, and, thus, the asymptotic theory reported here may not describe flows in geometries with sharp edges, such as the vicinity geometry~\citep{bandurin16,berdyugin19}.

\subsection{Fermi surface deformation}

The full Fermi surface deformation $h$ is reconstructed directly from the solution of the velocity profile and the Hall field as discussed in Sec.~\ref{sec:fulldistribution} and summarized in Table~\ref{tbl:hSummary}. Parametrizing the velocity in terms of the angle $\theta$ measured with respect to the positive $v_x$ axis [i.e., \mbox{$v_i = (\cos \theta, \sin \theta)$} and \mbox{$v_i^\perp = (\sin \theta, - \cos \theta)$}], the bulk deformation reads
\begin{subequations}
\begin{align}
h_B^{(0)} &= \delta \mu_B^{(0)} + 2 u^{(0)}_{B|y} \sin \theta , \label{eq:hB0channel} \\
h_B^{(2)} &= \delta \mu_B^{(2)} + 2 u^{(2)}_{B|y} \sin \theta 
- \sin(2\theta) \frac{\partial u^{(0)}_{B|y}}{\partial x} \nonumber \\
&+ \frac{1}{2} \frac{k_o}{1+(3k_o/r_c)^2} \sin(3\theta) \frac{\partial^2 u^{(0)}_{B|y}}{\partial x^2} \nonumber \\
&- \frac{3}{2} \frac{k_o^2/r_c}{1+(3k_o/r_c)^2} \cos(3\theta) \frac{\partial^2 u^{(0)}_{B|y}}{\partial x^2} , \label{eq:hB2channel}
\end{align}
\end{subequations}
where $u^{(0)}_{B|y}$ and $u^{(2)}_{B|y}$ are listed in Eqs.~\eqref{eq:uy0} and~\eqref{eq:uy2} and $\delta \mu_B^{(0)}$ and $\delta \mu_B^{(2)}$ follow by integrating the Hall field in Eqs.~\eqref{eq:mu0} and~\eqref{eq:mu2}. The structure of the bulk solution is directly apparent, with a hydrodynamic leading-order deformation $h_B^{(0)}$ that consists of a density and current deformation [the two terms in Eq.~\eqref{eq:hB0channel}] and a higher-order correction $h_B^{(2)}$ that, in addition to density and current deformations [the first two terms in Eq.~\eqref{eq:hB2channel}], contains a shear deformation [third term] as well as an odd-parity tomographic correction [fourth and fifth terms]. The tomographic boundary layer correction in Eq.~\eqref{eq:hT_form}, by contrast, contains all angular modes and can be directly evaluated in terms of the general harmonic coefficients $a_m$ and $b_m$ of the boundary-layer functions~$f_0$ and~$f_1$ reported in Sec.~\ref{sec:truncatedMomentExp}.

We illustrate this structure using the Fermi surface deformation in the channel. This deformation is shown in Fig.~\ref{fig:channel_FermiSurface} for $\lambda_m$ given by Eq.~\eqref{eq:lambda_m_model} with $p=0$ and exhibits a strong dependence on the tomographic effect and the strength of the applied magnetic field. Similar results are obtained for \mbox{$p=4$}. As $k_o$ increases in the absence of a magnetic field, the Fermi surface deformation exhibits a sharp feature for velocities parallel to the applied electric field [see Fig.~\ref{fig:channel_FermiSurface}(a)]. The sharp feature in the Fermi surface deformation is most prominent near the device edges (i.e., in the tomographic boundary layer).  When a magnetic field is applied, the sharp feature is suppressed once \mbox{$1/r_c\gtrsim 1/(3k_o)$} [see Fig.~\ref{fig:channel_FermiSurface}(b)]. This is attributed to the magnetic field, which both confines the tomographic boundary layer close to the boundary and suppresses the magnitude of tomographic phenomena, as was discussed previously.

\section{Summary and Outlook}\label{eq:conclusion}

In summary, we have derived an asymptotic theory describing steady two-dimensional tomographic electron flows in arbitrary smooth geometries. The central result of our work is a set of flow equations for the electrochemical potential~$\delta \mu$ and the velocity profile $u_i$, their boundary conditions, and corrections in the tomographic boundary layer, which are summarized in Table~\ref{tbl:BCs_tomographicCorrections}. We also report the corresponding microscopic Fermi surface deformation, summarized in Table~\ref{tbl:hSummary}. Our results were obtained using a systematic asymptotic expansion of the underlying kinetic equation that includes a generalized relaxation-time approximation and permits mode-dependent relaxation rates, where the expansion parameter is the small ratio of the even-parity mean free path to the length scale of the flow (the even-mode Knudsen number $k_e$). The existence of such an expansion is {\it a priori} not obvious, since the flow includes additional ballistic odd-mode degrees of freedom. The expansion is thus markedly distinct from analogous expansions of the classical kinetic equation~\cite{sone02,sone07} or conventional hydrodynamic theories with a single electronic relaxation time~\cite{benshachar25a,benshachar25b}.

Let us briefly comment on the effect of general boundary conditions. In the present work, we treat scattering from boundaries by a diffuse reflection condition, as appropriate for etched devices. An interesting extension for further study would be to include  extended boundary scattering models (such as a mixture of diffuse and specular reflection at the edges~\citep{holder19,raichev22}). Studies of nontomographic electron flows over an infinite flat edge~\citep{raichev22,kiselev19} and analogous studies of classical gas flows with nondiffuse reflection~\citep{onishi72,onishi74,onishi74b,onishi74c,benshachar25c} have found that nondiffuse scattering leads only to quantitative corrections, which manifest in the form of modified slip coefficients~\citep{raichev22,kiselev19,onishi72,onishi74,onishi74b,onishi74c,benshachar25c} and kinetic boundary layer corrections~\citep{onishi72,onishi74,onishi74b,onishi74c,benshachar25c}. For tomographic flows with general nondiffuse boundary conditions, we therefore expect that the asymptotic theory presented here will apply with an increased slip [i.e., with a change in the coefficients of Eq.~\eqref{eq:bulkBCslip}] and modified tomographic layer functions [which would follow from a solution of the tomographic equations with a modified boundary condition~\eqref{eq:hT2_eta0}].

The methodology presented here can (in principle) be extended to study thermal tomographic transport of electrons. This would require modeling the smeared Fermi surface at finite temperature, where the energy constraint is expected to differ from the electron number constraint, and, thus, give new equations for the energy current in the electron flow. This analysis would require the lifetimes of radial excitations of the Fermi surface, which were recently calculated (see, e.g., Ref.~\cite{nilsson24}).

Our framework offers both analytical insight into tomographic-flow effects and considerable versatility, since existing solvers of conventional Stokes-Ohm equations can easily be adapted to solve the presented equations that embody the asymptotic theory for tomographic flows. In addition, the comparison between the predictions of our asymptotic theory for flow in a channel and exact numerical solutions of the kinetic equation shows excellent agreement even at non-negligible values of the expansion parameter $\sqrt{k_e}$, which suggests that our description can be used as an alternative to expensive numerical solutions of the kinetic equation. Indeed, we expect the asymptotic theory to qualitatively describe the flow as long as the tomographic layers are confined near the boundaries and do not overlap one another, which requires \mbox{$\sqrt{k_e k_o}\ll 1$}. The theory reported here lays the foundations for future study of tomographic electron flows in other experimentally relevant geometries, and it would be interesting to combine it with numerical approaches to the Fermi-liquid equations~\citep{holder19,sulpizio19,estrada24,estrada24b,estrada25,raichev22,starkov25}. This is of particular importance for geometries in which hydrodynamic flows have been observed to exhibit novel phenomena, such as the rectification of electron flows in ``Tesla valve'' geometries~\citep{geurs20}, superballistic conductance of flows through constrictions~\citep{kumar17,stern22,krebs23} or through a lattice of obstructions~\citep{estrada24}.

We remark that recent measurements of the point-contact magnetoresistance  by~\textcite{ginzburg23} show a strong resistance increase at weak magnetic fields followed by a decrease in resistance at large magnetic fields, which is attributed to the tomographic effect (in particular, the backscattering of electrons as holes), with a predicted strong magnetoresponse in the tomographic-ballistic regime. Interestingly, the asymptotic theory applied to channel flow derived in the present work (which applies in the hydrodynamic-tomographic regime) exhibits the same qualitative features: a rapid increase in resistance at weak magnetic fields followed by a hydrodynamic suppression at large fields. However, the underlying mechanism differs between these scenarios: In the point-contact geometry the positive weak-field magnetoresponse arises since a  hole is backscattered into the constriction boundary rather than moving through the point contact, whereas in the asymptotic theory discussed here the positive magnetoresistance is due to the slip-velocity suppression at weak fields. Applying the framework of this paper to describe point-contact geometries would be an exciting prospect for further study. Finally, direct measurements of the Hall profile in a channel were reported by~\textcite{sulpizio19}, in which it was speculated that an observed enhanced curvature of the Hall field compared to the hydrodynamic prediction could be attributed to the tomographic effect. As discussed at the end of Sec.~\ref{eq:hallfieldchannel}, our results indicate that the tomographic effect does not account for this anomaly; rather, the curvature of the Hall field at the center of the channel is reduced by the tomographic effect. Moreover, we note that these experiments were conducted at a cyclotron radius comparable to the channel width, with \mbox{$1/r_c\approx1.3$} in our parametrization, at which value the asymptotic theory predicts a strong suppression of any tomographic phenomena.

A key prediction of our theory is that tomographic phenomena are suppressed for flows in any geometry when the cyclotron radius is comparable to or smaller than the odd-mode mean free path, $R_c\lesssim \ell_o$. This suggests a simple \textit{in situ} test for tomographic effects in confined geometries---analogous to the magnetic-field-induced suppression of viscous effects described in Ref.~\citep{zeng24}---which can be readily implemented in present-day experiments.

\begin{acknowledgments}
This work is supported by Vetenskapsr\aa det (Grants No.~2020-04239 and No.~2024-04485), the Olle Engkvist Foundation  (Grant No.~233-0339), the Knut and Alice Wallenberg Foundation (Grant No.~KAW 2024.0129), and Nordita.  We acknowledge support from the University of Melbourne’s Research Computing Services and the Petascale Campus Initiative. N. B.-S.~acknowledges support from the Australian Government Research Training Program Scholarship.
\end{acknowledgments}

\appendix

\section{LINEARIZING THE ELECTRON BOLTZMANN EQUATION} \label{app:linearization}

In this appendix, we summarize the linearization procedure of the Boltzmann equation for conduction electrons. This produces the linearized scaled Boltzmann equation in Eq.~\eqref{eq:BTE}. The starting point is the dimensional Boltzmann equation for the full distribution function $f$:%\johannes{Should it be $e \to e/m^*$?}
\begin{align}
    v_i \frac{\partial f}{\partial x_i} + \frac{e}{m^*}\biggl(\mathcal{E}_i -\frac{\partial \tilde{\Phi}}{\partial x_i}+ \varepsilon_{ijk}v_j\mathcal{B}_k\biggr) \frac{\partial f}{\partial v_i} && \nonumber \\[1ex]
    = I_e[f] + I_o[f] +I_\text{MR}[f] &,\label{eq:dimensionalBTE}
\end{align}
where $x_i$ and $v_i$ are (dimensional) position and velocity coordinates, respectively, $\tilde{\Phi}$ is the dimensional induced electric potential, and ${\cal B}_i$ and ${\cal E}_i$ are the external (dimensional) magnetic and electric fields, respectively. The left-hand side in Eq.~\eqref{eq:dimensionalBTE} is the streaming term, and the right-hand side is the collision term. 
For the latter, we adopt a generalized relaxation-time approximation that accounts for momentum-conserving collisions (described by the nonlinear operators $I_{e}$ and $I_o$ for the relaxation of even- and odd-parity components of the distribution function, respectively) as well as residual momentum-relaxing collisions (described by the nonlinear operator $I_\text{MR}$) with a relaxation time $\tau_\text{MR}$. This relaxation time is related to the collisional mean free path via \mbox{$\ell_\text{MR}=v_F \tau_\text{MR}$}. The momentum-conserving and momentum-relaxing collisions relax the distribution function to  local moving and stationary equilibria, respectively:
\begin{subequations}
\begin{align}
    f_\text{MC} = & \left[1+\exp\left(\frac{\frac{1}{2}m^*(v_i-\bar{v}_i)^2-\mu}{k_B T}\right)\right]^{-1}, \\
    f_\text{MR} = & \left[1+\exp\left(\frac{\frac{1}{2}m^*v_i^2-\mu}{k_B T}\right)\right]^{-1},
\end{align}\label{eq:localEquil}%
\end{subequations}
where $T$ is the local temperature, $\mu$ is the local chemical potential, and $\bar{v}_i$ is the local drift velocity. We assume a parabolic dispersion relation throughout. In the following, we expand Eq.~\eqref{eq:dimensionalBTE} for low temperatures relative to the Fermi temperature and for weak driving electric fields, for which the above approximations for the collision operator on the right-hand side in Eq.~\eqref{eq:dimensionalBTE} are applicable.

To linearize Eq.~\eqref{eq:dimensionalBTE}, we first scale the velocity and spatial coordinates by $v_F$ and $L$, respectively. The applied electric and magnetic fields are scaled by their respective magnitudes $\mathcal{E}$ and $\mathcal{B}$, and the induced electric potential by $\mathcal{E}L$:
\begin{align}
    \mathcal{E}_i = \mathcal{E} E_i, \qquad \mathcal{B}_i = \mathcal{B} B_i, \qquad \tilde{\Phi} = \mathcal{E}L\Phi .
\end{align}
We then introduce the electron Mach number
\begin{align}
    \text{Ma} = \frac{U}{v_F}, \label{eq:MaDefn}
\end{align}
where $U=e\mathcal{E}L^2/(m^* v_F^2\tau_e)$ is the hydrodynamic velocity scale. The Boltzmann equation is linearized by expanding the distribution function and each of the macroscopic variables for a small Mach number, followed by an expansion for temperature small relative to the Fermi temperature:
\begin{align}
    \text{Ma} \ll 1, \qquad \frac{T}{T_F} \ll 1 .
\end{align}
The former expansion is applicable for weak driving electric fields, and the latter is applicable to degenerate systems and confines all excitations to the Fermi surface, which is assumed isotropic. The expansions for the distribution function and electrochemical potential then read
\begin{subequations}
\begin{align}
    f &= f_0 + \text{Ma}\, E_F \biggl(- \frac{\partial f_0}{\partial E}\biggr) \left(h+2k_e \Phi\right) \nonumber \\
    &\qquad + {\it O}\biggl(\text{Ma}^2,\biggr(\frac{T}{T_F}\biggr)^2\biggr), \\
    \mu &= E_F \biggl[1+ \text{Ma} \left(\delta\mu+2k_e \Phi\right)+{\it O}\biggl(\text{Ma}^2,\biggr(\frac{T}{T_F}\biggr)^2\biggr) \biggr], 
\end{align}\label{eq:smallMaExpansion}%
\end{subequations}
respectively, where $f_0$ is the global equilibrium distribution, \mbox{$E=m^*v^2/2$} is the energy of the electrons, and \mbox{$E_F= m^*v_F^2/2$} is the Fermi energy. Furthermore, $h$ and $\delta \mu$ are the scaled linear perturbations of the distribution function and electrochemical potential, respectively. The dimensionless bulk velocity $u_i$ is defined via \mbox{$\bar{v}_i = v_F \text{Ma}\,u_i$}, which follows directly from the definition of the Mach number in Eq.~\eqref{eq:MaDefn}. The macroscopic variables $\delta\mu$ and $u_i$ are given by the zeroth and first moments, respectively, of the distribution function as reported in  the main text [Eqs.~\eqref{eq:momentsdensity} and~\eqref{eq:momentsvelocity}]. Substituting the expansion in Eq.~\eqref{eq:smallMaExpansion} into Eqs.~\eqref{eq:dimensionalBTE} and~\eqref{eq:localEquil} and collecting terms linear in $\text{Ma}$ gives the linearized Boltzmann equation~\eqref{eq:BTE} reported in the main text, where we have further assumed that even-mode relaxation is described by a single relaxation time $\tau_e=\ell_e/v_F$.

\section{COORDINATE FORM OF STRESS-TENSOR DERIVATIVES}\label{app:derivatives}

This appendix collects some derivative identities that are used to obtain the coordinate representation of the boundary condition~\eqref{eq:hT2_eta0}. Similar identities have been reported in the classical gas literature; see, for example, Refs.~\cite{sone69,sone71,nassios12}. First, to express the higher-moment even-parity correction in coordinate form, we note
\begin{subequations}
\begin{align}
\biggl[ t_i t_j \frac{\partial u_{B|i}^{(0)}}{\partial x_j} \biggr]_S &= 0 , \\
\biggl[ n_i n_j \frac{\partial u_{B|i}^{(0)}}{\partial x_j} \biggr]_S &= 0 .
\end{align}
\end{subequations}
The first identity follows from the no-slip boundary condition \mbox{$t_i u_{B|i}^{(0)}=0$}, which holds identically along the boundary, and the second one follows from the continuity equation and the no-slip condition. Since these equations hold identically along the boundary, they imply the additional identities
\begin{subequations}
\begin{align}
    \biggl[ t_k \frac{\partial}{\partial x_k} \biggl(t_i t_j \frac{\partial u_{B|i}^{(0)}}{\partial x_j}\biggr) \biggr]_S &= 0 , \\
    \biggl[ t_k \frac{\partial}{\partial x_k} \biggl(n_i n_j \frac{\partial u_{B|i}^{(0)}}{\partial x_j}\biggr) \biggr]_S &= 0 .
\end{align}
\end{subequations}
To rewrite the higher-order odd-parity contributions, we need the additional identity
\begin{align}
\biggl[ n_i t_j t_k \frac{\partial^2 u_{B|i}^{(0)}}{\partial x_j \partial x_k} \biggr]_S &= 0 , \label{eq:nttSH}
\end{align}
which again follows from the no-penetration boundary condition that holds identically along the boundary. In addition, we have
\begin{align}
\biggl[ n_i n_j n_k \frac{\partial^2 u_{B|i}^{(0)}}{\partial x_j \partial x_k} \biggr]_S &= - \frac{1}{2} \biggl[ n_i n_j n_k \frac{\partial S_{B|ij}^{(0)}}{\partial x_k} \biggr]_S , \label{eq:nnnSH}
\end{align}
where the rate-of-strain tensor is defined in Eq.~\eqref{eq:stresstensor}. Furthermore,
\begin{align}
\biggl[ n_i t_j n_k \frac{\partial^2 u_{B|i}^{(0)}}{\partial x_j \partial x_k} \biggr]_S &= \kappa \Bigl[n_i t_j S_{B|ij}^{(0)} \Bigr]_S .
\end{align}
We use this relation to obtain the following identity:
\begin{align}
&\biggl[ t_i n_j n_k \frac{\partial^2 u_{B|i}^{(0)}}{\partial x_j \partial x_k} \biggr]_S \nonumber \\[1ex]
&\quad =\biggl[ - t_i n_j n_k \frac{\partial S_{B|ij}^{(0)}}{\partial x_k} \biggr]_S - \kappa \Bigl[ n_i t_j S_{B|ij}^{(0)} \Bigr]_S.  \label{eq:tnnSH}
\end{align}
In addition, using the incompressibility condition gives
\begin{align}
\biggl[ t_i n_j t_k \frac{\partial^2 u_{B|i}^{(0)}}{\partial x_j \partial x_k} \biggr]_S &= \frac{1}{2} \biggl[ n_i n_j n_k \frac{\partial S_{B|ij}^{(0)}}{\partial x_k} \biggr]_S .
\end{align}
Finally,
\begin{align}
\biggl[ t_i t_j t_k \frac{\partial^2 u_{B|i}^{(0)}}{\partial x_j \partial x_k} \biggr]_S &= - \kappa \Bigl[ n_i t_j S_{B|ij}^{(0)} \Bigr]_S .  \label{eq:tttSH}
\end{align}
We also need the Laplace identities
\begin{subequations}
\begin{align}
\biggl[ n_i \frac{\partial^2 u_{B|i}^{(0)}}{\partial x_j^2} \biggr]_S &= - \frac{1}{2} \biggl[ n_i n_j n_k \frac{\partial S_{B|ij}^{(0)}}{\partial x_k} \biggr]_S , \\
\biggl[ t_i \frac{\partial^2 u_{B|i}^{(0)}}{\partial x_j^2} \biggr]_S &= - 2 \kappa \Bigl[ n_i t_j S_{B|ij}^{(0)} \Bigr]_S - \biggl[ t_i n_j n_k \frac{\partial S_{B|ij}^{(0)}}{\partial x_k} \biggr]_S ,
\end{align}
\end{subequations}
where we used Eqs.~\eqref{eq:nnnSH} and~\eqref{eq:nttSH} for the first identity and Eqs.~\eqref{eq:tnnSH} and~\eqref{eq:tttSH} for the second identity.

Taken together, we obtain the coordinate representation
\begin{subequations}
\begin{align}
&\biggl[ v_i v_j \frac{\partial u_{B|j}^{(0)}}{\partial x_i} \biggr]_S = - \frac{1}{2} \sin 2\theta \, \Bigl[ n_i t_j S_{B|ij}^{(0)} \Bigr]_S  ,\\[1ex]
&\biggl[v_i v_j v_k - \frac{1}{4} \delta_{ik} v_j\biggr] \biggl[\frac{\partial^2 u_{B|j}^{(0)}}{\partial x_i \partial x_k} \biggr]_S \nonumber\\
&= 
- \frac{\kappa}{2} \cos (3 \theta) \, \Bigl[n_i t_j S_{B|ij}^{(0)} \Bigr]_S + \frac{\cos (3 \theta)}{4} \, \biggl[t_i n_j n_k \frac{\partial S_{B|ij}^{(0)}}{\partial x_k} \biggr]_S \nonumber\\
&\quad + \frac{3}{8} \sin (3 \theta) \, \biggl[ n_i n_j n_k \frac{\partial S_{B|ij}^{(0)}}{\partial x_k} \biggr]_S ,\\[1.5ex]
&\biggl[v_i^\perp v_j^\perp v_k^\perp - \frac{1}{4} \delta_{ik} v_j^\perp\biggr] \biggl[\frac{\partial^2 u_{B|j}^{(0)}}{\partial x_i \partial x_k} \biggr]_S \nonumber\\
&= 
\frac{\kappa}{2} \sin (3 \theta) \, \Bigl[n_i t_j S_{B|ij}^{(0)} \Bigr]_S - \frac{1}{4} \sin (3 \theta) \, \biggl[t_i n_j n_k \frac{\partial S_{B|ij}^{(0)}}{\partial x_k} \biggr]_S \nonumber\\
&\quad + \frac{3}{8} \cos (3 \theta) \, \biggl[ n_i n_j n_k \frac{\partial S_{B|ij}^{(0)}}{\partial x_k} \biggr]_S .
\end{align}
\end{subequations}

\section{TOMOGRAPHIC LAYER FUNCTIONS FOR MODE-DEPENDENT SCATTERING}\label{app:p4_tomoFun}

\begin{figure*}
    \centering
    \includegraphics[width=0.9\linewidth]{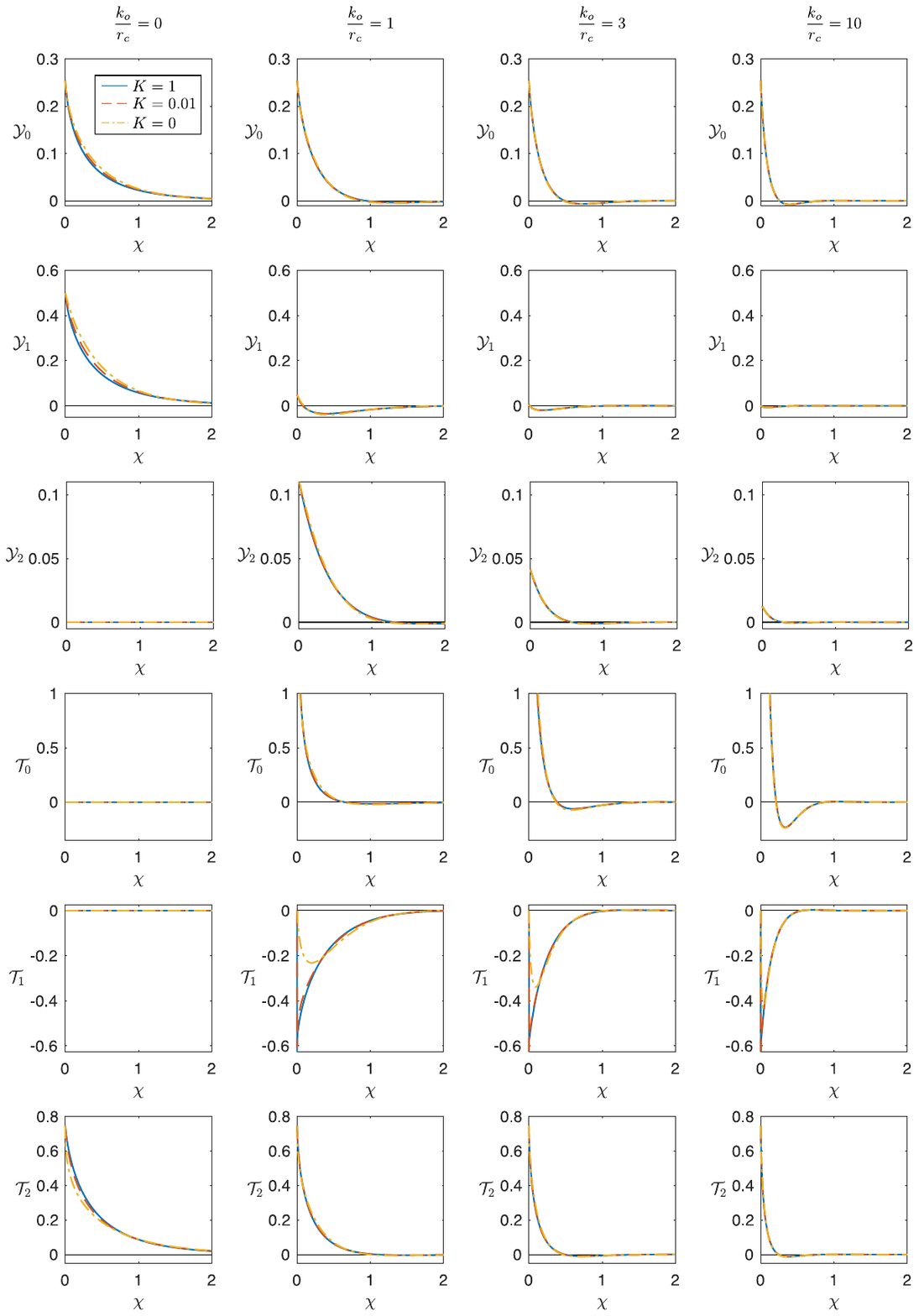}
    \caption{Tomographic layer functions $\mathcal{Y}_0, \mathcal{Y}_1, \mathcal{Y}_2, \mathcal{T}_0, \mathcal{T}_1$, and $\mathcal{T}_2$ (top to bottom), calculated for $k_o/r_c=0,1,3,10$ (from left to right) for mode-dependent odd-mode relaxation rates given by Eq.~\eqref{eq:lambda_m_model} with $p=4$ and $K=1$ (blue solid line), $K=0.01$ (red dashed line), and $K=0$ (yellow dot-dashed line).}
    \label{fig:tomo_functions_p4}
\end{figure*}

In this appendix, we present solutions for the tomographic layer functions for mode-dependent relaxation rates corresponding to Eq.~\eqref{eq:lambda_m_model}. The tomographic layer functions now depend on both $k_o/r_c$ (as for the constant odd-mode relaxation model) and~$K$ and~$p$. We note that, for \mbox{$K=0$}, Eq.~\eqref{eq:lambda_m_model} recovers the collision model of Refs.~\citep{ledwith17,ledwith19}, in which the relaxation rate of the $m$th odd mode scales exactly with $m^p$. For \mbox{$p=0$}, Eq.~\eqref{eq:lambda_m_model} recovers the constant odd-mode relaxation rate of Ref.~\citep{hofmann22}. Solutions for the tomographic layer functions $\mathcal{Y}_0, \mathcal{Y}_1, \mathcal{Y}_2, \mathcal{T}_0, \mathcal{T}_1$, and $\mathcal{T}_2$ are shown in Fig.~\ref{fig:tomo_functions_p4} for \mbox{$p=4$}, \mbox{$K\in\{1,0.01,0\}$}, and \mbox{$k_o/r_c\in\{0,1,3,10\}$}. All tomographic layer functions exhibit a weak dependence on $K$, except for $\mathcal{T}_1$, whose gradient near \mbox{$\chi=0$} is seen to be of opposite sign for \mbox{$K=0$} and \mbox{$K=1$}.

\section{NUMERICAL SOLUTION FOR CHANNEL FLOW} \label{app:numSol}

This appendix summarizes the numerical solutions to Eq.~\eqref{eq:BTE} for flow in a channel, which we use in the main text to compare with the result of the asymptotic expansion. Further details are given in Ref.~\citep{benshachar25b}. Equation~\eqref{eq:BTE} is discretized with $N_x$ and $N_\theta$ points in the spatial and velocity coordinates, respectively, where the spatial points are concentrated near the boundaries \mbox{$x=\pm 1/2$}. Derivatives are then approximated at each mesh point using a finite-volume method, and moments of the distribution function are approximated using the trapezoidal rule. A time derivative is added to the left-hand side of Eq.~\eqref{eq:BTE}; the resulting kinetic equation is discretized using the second-order Runge-Kutta method and iterated from global equilibrium until a steady state is achieved. The number of spatial and velocity points is then increased until the solution changes by less than 1\% with further mesh refinement. A final mesh of \mbox{$N_x=300$} and \mbox{$N_\theta=200$} is used.

\section{COMPARISON OF THE ASYMPTOTIC AND NUMERICAL SOLUTIONS FOR CHANNEL FLOW}\label{app:ke_convergence} 

\begin{figure}
    \centering
    \includegraphics{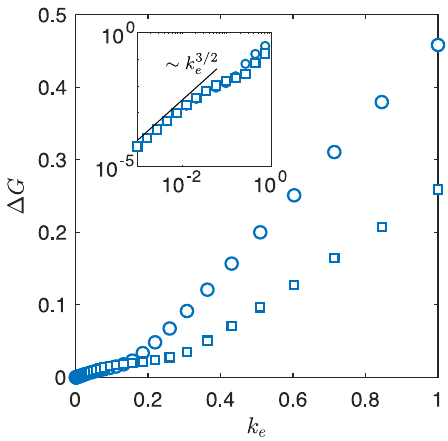}
    \caption{
    Relative error in the channel conductance between the asymptotic and direct numerical solutions. Parameters are \mbox{$G\to\infty$} and  \mbox{$k_o=2$} with an odd-mode relaxation rate given by Eq.~\eqref{eq:lambda_m_model} with \mbox{$p=4$}, and \mbox{$1/r_c=0$} (circles) and \mbox{$1/r_c=0.1$} (squares). The inset shows the same data on a double-logarithmic scale, along with the scaling $\sim k_e^{3/2}$ as a black line.}
    \label{fig:ke_scaling}
\end{figure}

In this appendix, we compare the asymptotic solution for channel flow derived in Sec.~\ref{sec:channel_flow} with direct numerical solutions, which demonstrates that the matched asymptotic expansion considered in this work is quantitatively accurate over a large range of Knudsen numbers~$k_e$. To this end, Fig.~\ref{fig:ke_scaling} shows as a function of~$k_e$ the relative error in the channel conductance, i.e.,
\begin{align}
    \Delta G = \frac{\int_{-1/2}^{1/2} dx \, [u_y^{\rm a}(x)-u_y^{\rm n}(x) ]}{\int_{-1/2}^{1/2} dx \, u_y^{\rm n}(x)},
\end{align}
where $u_y^{\rm a}$ is the velocity profile predicted by the asymptotic theory discussed in Sec.~\ref{sec:channel_flow} and $u_y^{\rm n}$ is the velocity profile obtained from a direct numerical solution of the linearized Boltzmann equation as described in Appendix~\ref{app:numSol}. We choose an odd-mode Knudsen number \mbox{$k_o=2$} with a mode-dependent odd-mode relaxation rate given by Eq.~\eqref{eq:lambda_m_model} and negligible impurity scattering \mbox{$G\to\infty$}. The figure shows the error without a magnetic field (circles) and in the presence of a magnetic field with \mbox{$1/r_c=0.1$} (squares). As is apparent from the figure, even for relatively large Knudsen numbers, such as~\mbox{$k_e= 0.4$}, the error between the asymptotic prediction and the full numerical results is small, less than approximately~\mbox{$12\%$}, and it is less than~\mbox{$3\%$} for \mbox{$k_e\lesssim 0.2$}. A finite magnetic field, which suppresses higher angular modes, is seen to further suppress the error and extend the regime of applicability of the asymptotic theory to larger Knudsen numbers. Note, moreover, that at small values of $k_e$ the error in the asymptotic theory scales as $k_e^{3/2}$ (inset figure), as expected from the truncation of the expansion in Eq.~\eqref{eq:keExpansion}.

\FloatBarrier

\bibliography{bib_knudsen}

\end{document}